\documentclass{elsart}
\usepackage{natbib}
\usepackage{graphicx}
\usepackage{amsmath, amssymb}
\usepackage{cite, citesort}

\begin{document}

\begin{frontmatter}


\title{Exploitation dynamics of fish stocks}
%
%
\author{Hiro-Sato Niwa\thanksref{phone}}
\thanks[phone]{Tel.: +81-479-44-5953; fax: +81-479-44-1875.\\
\hspace*{3mm} {\it E-mail address:} Hiro.S.Niwa@fra.affrc.go.jp (H.-S. Niwa).}

\address{Behavioral Ecology Section,
National Research Institute of Fisheries Engineering,
7620-7 Hasaki, Kamisu, Ibaraki 314-0408, Japan}

\begin{abstract}
I address the question of the fluctuations in fishery landings.
Using the fishery statistics time-series collected by the Food and
 Agriculture Organization of the United Nations since the early 1950s,
I here analyze fishing activities and find two scaling features of
 capture fisheries production:
(i) the standard deviation of growth rate of the domestically landed
 catches decays as a power-law function of country landings with an
 exponent of value $0.15$;
(ii) the average number of fishers in a country scales to the $0.7$
 power of country landings.
I show how these socio-ecological patterns may be related, yielding a
 scaling relation between these exponents.
The predicted scaling relation implies that
the width of the annual per capita growth-rate distribution scales to
 the $0.2$ power of country landings, i.e.
 annual fluctuations in per capita landed catches increase with
 increased per capita catches in highly producing countries.
Beside the scaling behavior, I report that
 fluctuations in the annual domestic landings have increased in the last
 30 years,
 while the mean of the annual growth rate declined significantly after
 1972.
\end{abstract}


\end{frontmatter}

%

\section{Introduction}
Fluctuations in fishery landings have been a collective human concern
\citep{Botsford-etal97,Pauly-etal02}.
Variation in supply (stock size) is a central problem in determining
the optimal organization of the fishing industry and in how governments
can best regulate fishing effort.
Since the industrialization of fishing, fisheries scientists have
been subject to intense political pressures:
``Determine the causes of fluctuations in fish catches.''
For example,
in the mid-1800s the Norwegian parliament asked zoologists
Michael Sars and George Ossian Sars
to investigate the biology of Norwegian fisheries and to answer why
the cod catches from the Lofoten Islands in Northern Norway fluctuated
so greatly \citep{Smith94}.

Nonlinear dynamics have been applied to understanding the ups and downs
of populations.
In the mid-1920s,
an Italian biologist Umberto D'Ancona
found the increase in predator fish and decrease in prey fish of various
species in the Adriatic Sea during the World War I period from a
statistical study of the numbers of each species sold on the fish
markets of three ports, Fiume, Trieste, and Venice
\citep{D'Ancona26,D'Ancona54}.
A famous Italian mathematician Vito Volterra,
asked by D'Ancona whether
he could come up with a mathematical model of predator-prey population
dynamics,
developed the first and simplest model of predator-prey interactions
\citep{Volterra26}.
Because an American mathematical biologist Alfred J. Lotka produced
the two-species interaction model independently and at about the same
time \citep{Lotka25},
it is called Lotka-Volterra model.
The nonlinear model predicts a cyclical relationship between predator
and prey numbers.

Thus, the jagged oscillation in populations was nothing new to
ecologist, but before the 1970s, they put most of the patterns down to
the stochastic, unaccountable effects:
year-to year fluctuations in fish stocks are usually attributed to
variability in recruitment, competition, predation, changes in fishing
effort, and other sources of so-called environmental noise.
\citet{May74,May76} discovered that a density-dependent (i.e. nonlinear),
deterministic model of ecology could produce complex, random-looking
patterns ``chaos''.
\citet{Conklin-Kolberg94}
researched the effects of market demand on
halibut populations of Pacific Ocean fisheries,
and provided a model of chaotic dynamics for the Pacific halibut
fishery.
\citet{Weisbuch-etal97} studied the influence of biological and
socio-economic mechanisms on the amplitude of resource depletion, and
showed not only that
the depletion is most often due to inertial factors in the dynamics
of exploitation, i.e. capital and labor overshoot,
but also that capital and labor inertia results into dangerous
oscillations of the resource level, eventually finishing with a fishery
crash.

A central question in ecology for last decades has been how to quantify
the relative importance of stochastic and nonlinear factors for
fluctuations in population size
\citep{Saether-etal00,Turchin03,Hsieh-etal05}.
Characterizing the fluctuation patterns is the major challenge for
time series analysis in fisheries and for modeling the process of fish
production.
One major approach to dealing with ecological complexity
is to reduce the system to one or a few species:
a number of detailed analysis of specific systems have captured the
patterns of fluctuations in fish stocks and catches,
including noise-driven fluctuations and chaos
\citep{Wilson-etal91,Matsuda-etal91,Matsuda-etal92,Takeuchi-etal92,Higgins-etal97,Bjornstad-etal99,Dixon-etal99,Bjornstad-Grenfell01}.

Inspired by work of \citet{Keitt-Stanley98} and \citet{Keitt-etal02},
I here explore an alternative approach to
socio-ecological system dynamics, i.e.
population dynamics within which fisheries are embedded.
From the macro-ecological viewpoint,
I analyze the time-series of annual fishery landings (all commercial
species items fished inclusive) of 244 countries in total during the
period 1950--2002,
after excluding non-fish vertebrates such as whales.
The data on annual domestically-landed catch are
taken from Fishstat Plus \citep{FaoFishStat+},
a set of statistical database compiled by the Food and Agriculture
Organization (FAO) of the United Nations.
FAO is the only one institution that maintains global fisheries
statistics.
The country-structured time-series of landed catch is a complex source
of information.
Annual landed catch is the aggregate of a large number of complex
interactions between fishers and the stock.
Some aspects of landed catch are biological phenomena (biomass of fish
vulnerable to the gear) and technical efficiencies (use of machinery
power and modern equipments such as echo sounders and fish detection
devices);
others are economic (market price of fish and the cost of applying
effort), legal (regulations applied to fishing effort) and some are
behavioral (skill of the fisher targeting the stock and accurate
completion of catch records).
Besides the fluctuations in fishery landings,
I investigate the employment statistics based on the FAO data on the
number of commercial and subsistence fishers \citep{NoFishers,SOFIA02}.
I report two power-law relationships between the following quantities
for the capture fisheries sector:
\begin{itemize}
 \setlength{\itemsep}{10pt}
 \item[{(i)}] the variability of landings time-series and the country
		 landings
 \item[{(ii)}] the number of fishers in a country and the country
		 landings
\end{itemize}
In order to understand the scaling structure for the stock exploitation,
a relation between these power-law exponents is hypothesized,
such as the Widom scaling relation \citep{Widom65a,Widom65b}.

In the paper, symbols $x$ and $n$ are used for the annual
domestically-landed catch and the number of fishers in countries,
respectively.

\section{World fisheries trend}
Fisheries have recently become a topic for the media with global
audiences, as well as for the scientific community
[e.g. \citet{Beddington-Kirkwood05,Ehrhardt05,Symes05}],
with a focus on the wretched history, the poor current practice, and
the future scenarios.
For instance,
the New Orleans {Times-Picayune}'s eight-part 1996 series by 
\citet{McQuaid-etal96}
on ``Oceans of Trouble'' won the Pulitzer Prize for public service
in 1997,
in reporting on threats to the world's fisheries.

In 1950, the newly founded Food and Agriculture Organization of
the United Nations began collection of global statistics.
Fisheries in the early 1950s were at the onset of a period of
extremely rapid growth.
Everywhere that industrial-scale fishing (mainly trawling, but also
purse seining and long-lining) was introduced, it competed with
small-scale, or artisanal fisheries.
Throughout the 1950s and 1960s, this huge increase of global fishing
effort led to an increase in catches so rapid that their trend
exceeded human population growth, encouraging an entire generation of
fisheries managers and politicians to believe that launching more boats
would automatically lead to higher catches \citep{Pauly-etal02}.

In 2002, fisheries provided direct employment to about 38 million people
and accounted for more than 15 percent of the total human consumption of animal
protein \citep{SOFIA02}.
Globally, first-sale capture-fishery revenues produced about
78 billion US dollars.
Total world trade of fish products increased to an export
value of 58.2 billion US dollars, continuing the last decade's
underlying 4.5 percent annual growth.
The world's population has been increasing more quickly than the total
food fish supply; as a result the average per capita fish supply has
declined.
Total production from marine and inland capture
fisheries for the world
reached 93.2 million tonnes
(84.5 million tonnes marine and 8.7 million tonnes inland)
[excluding China, 76.6 million tonnes from capture fisheries
(70.1 million tonnes marine and 6.5 million tonnes inland)].

\begin{figure}[p]
 \centering
 \includegraphics[width=6.5cm]{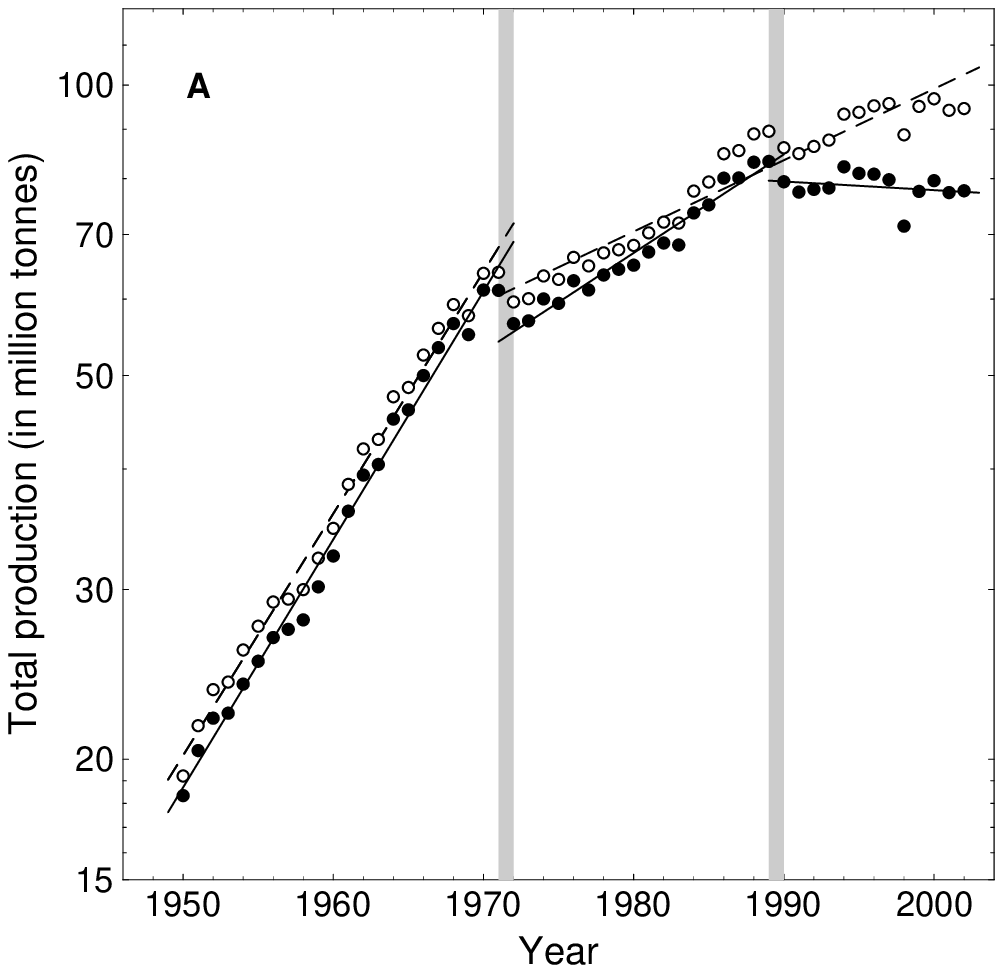}
 \hspace{0.5cm}
 \includegraphics[width=6.5cm]{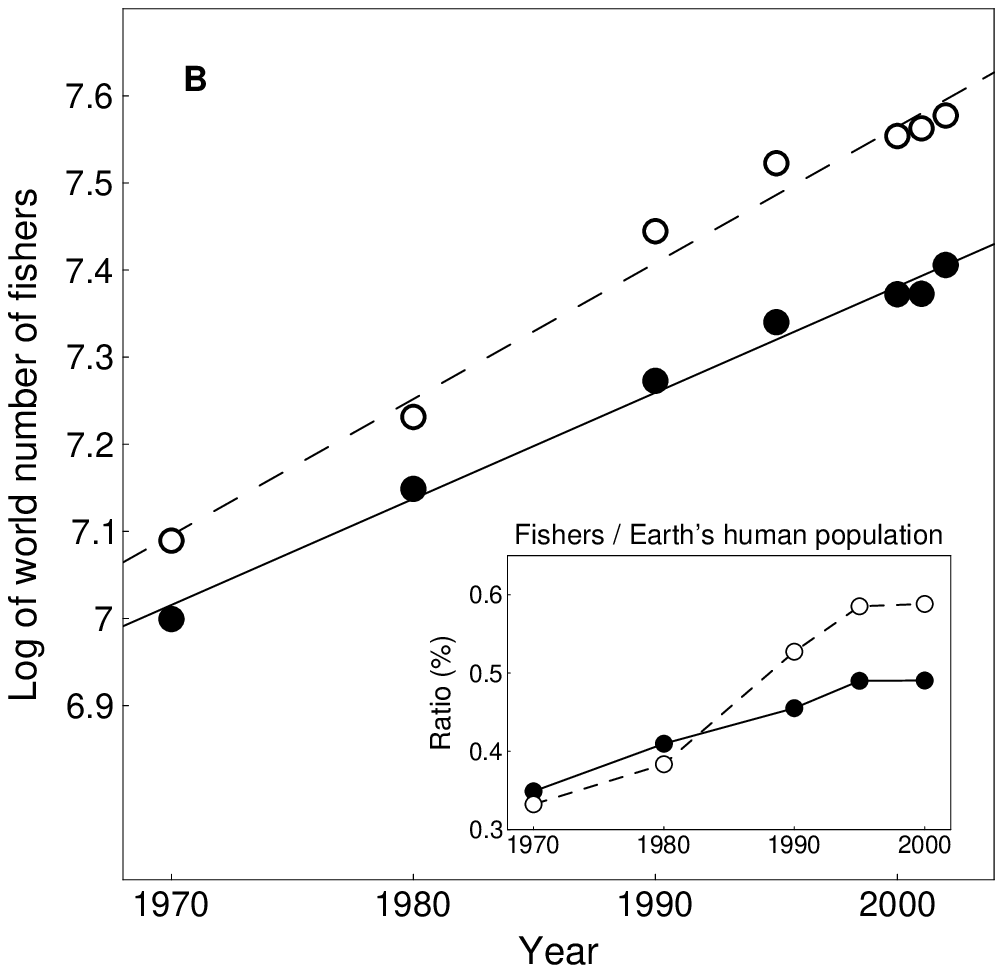}
 \caption{}
 \label{fig:1}
\end{figure}

Figure~\ref{fig:1}{\sf A} shows the trend of the global harvests with
capture fisheries,
demonstrating that the global capture production has been exponentially
growing.
The annual growth rate of global catches
(defined as $R \equiv \ln [X(t+1)/X(t)]$,
where $X(t)$ and $X(t+1)$ are the total capture production in the years
$t$ and $t+1$)
changed discontinuously at the critical time 1972,
correspondent to the collapse of the Peruvian anchoveta fishery,
signaling an abrupt transition of bio-economic regime.
The fisheries exploitation has expanded in two-exponential mode, and
possibly reached a plateau with no transient period
contrary to a logistic growth.

In line with the increase in fisheries production, over the last three
decades employment in fisheries and aquaculture has continued to
increase in many countries.
The world number of fishers and fish farmers has been exponentially
(i.e. in a Malthusian mode)
growing at annual rate of 3.6 percent
(excluding China, 2.8 percent)
for the period 1970--2002,
depicted in Fig.\ref{fig:1}{\sf B}.
The world fishers population has been growing at a faster rate than the
earth's human population 
(see inserted figure of Fig.\ref{fig:1}{\sf B}).
Since 1970 to 2000,
the ratio of fishers and fish farmers to the whole population
has increased from 0.332 to 0.588 percent
(excluding China, 0.349 to 0.490 percent).

Since there is the possibility of China's misreporting \citep{WatsonPauly01},
for further analysis, I use the data with the exclusion of China's
reported catches
(including China does not change the results).

\section{Scaling of country growth rate}
In Fig.\ref{fig:2}, shown are the probability distributions
of the annual domestic landings~({\sf A}) and the number of people
actively fishing and fish farming in countries~({\sf B}),
both which are consistent with a log-normal distribution
\begin{equation}
 P(x)
  =
  \frac 1{x\sqrt{2\pi s_x^2}}
  \exp
  \left[-\frac{\ln^2 (x/\hat{x})}{2s_x^2} \right],
  \label{eqn:lognormal-landings}
\end{equation}
where $\hat{x}$ and $s_x$ denote a geometric mean and a geometric standard
deviation of the distribution, respectively.
The log-normal relationship remained unchanged for the time
period studied.
This log-normal distribution of the country landings is suggestive of
Gibrat's law on market structure \citep{Gibrat31}.
It is interesting to note that,
contrary to city sizes (resident population) or US firm sizes (employees
and revenue),
the country population engaged in fishing and fish farming does not
reduce to the Zipf's power-law distribution characterized by the absence
of a characteristic size \citep{Zipf47,Axtell01}.
However, the conditional probability distributions of fishers population
for countries with the same landings show ``fat tails'' with
characteristic scales
(for details, consult \S \ref{section:stretched-exponentials}).

In the limit of small annual changes in $x$, we can define the relative
growth rate as
\begin{equation}
 r
 \equiv
 \ln\left[\frac{x(t+1)}{x(t)}\right]
\approx
\frac{x(t+1)-x(t)}{x(t)},
\end{equation}
where $x(t+1)$ and $x(t)$ are the annual landings of a country in two
consecutive years.
I calculate the conditional distribution $p(r|x)$ of annual
growth rate for countries with landings $x$.
The distribution $p(r|x)$ of the growth rate from 1950 to 1971 is
shown in Fig.\ref{fig:3}{\sf A}, depicting a simple tent-shaped form on
a semi-logarithmic scale.
The distribution is not Gaussian, as expected from the Gibrat model, but
rather is exponential:
\begin{equation}
 p(r|x)
  =
  \frac 1{\sqrt{2}\sigma_r (x)}
  \exp \left(
	-\frac{\sqrt{2} \left| r-\overline{r}(x)\right|}{\sigma_r (x)}
      \right),
  \label{eqn:laplace}
\end{equation}
where $\overline{r}(x)$ and $\sigma_r (x)$ are the average growth rate
and the standard deviation of $r$ for countries with landings equal to
$x$, respectively.
An implication of this result is that the distribution of the growth
rate has much broader tails than would be expected for a Gaussian
distribution.
Notice that the fluctuation $\sigma_r$ in growth rate $r$ is
independent of $x$.

In Fig.\ref{fig:3}{\sf B}, shown is the empirical conditional
probability density of $r$ for countries from 1972 to 2002 with
approximately the same landings,
suggesting that $p(r|x)$ has the same functional form,
consistent with the Laplace distribution [Eq.(\ref{eqn:laplace})],
with different widths $\sigma_r (x)$, for all $x$.
In these years, contrary to the period of 1950--1971,
we expect that the statistical properties of the growth rate $r$ depend
on landed catch $x$:
the growth-rate distribution depends on the initial value of $x$.

The study of the fluctuations has been shown to give important
information regarding the underlying processes responsible for the
observed macroscopic behavior.
I therefore analyze the fluctuations in the growth rate of the country
landings.
While different plots for the values of landings fall onto one
another for the 1950--1971 period,
there is clear dependence of $\sigma_r$ on the value of landed
catch for both the periods of 1972--1989 and 1990--2002.
Figure~\ref{fig:4}{\sf A} shows that for the periods 1972--2002,
$\sigma_r (x)$ statistically decreases with increasing $x$.
The width of the growth-rate distribution, $\sigma_r (x)$, scales over
three order of magnitude, from landings about $10^{3}$ tonnes up to
landings about $10^{6}$ tonnes, as a power law:
\begin{equation}
 \sigma_r (x) = A x^{-\beta}
  \label{eqn:scaling-growth-rate}
\end{equation}
where $A \approx 0.27$ and $\beta \approx 0.15$ for the both
investigated periods
(landed value of catch unit in thousand tonnes).

What is remarkable about Eqs.(\ref{eqn:laplace}) and
(\ref{eqn:scaling-growth-rate}) is that they govern the growth rates of
a diverse set of countries.
Fishery labor productivity and capital intensity vary widely among
countries.
They range not only in their amount of landings but also in what and
where they catch:
the harvest included in the database is classified according to
more than 1000 commercial species items,
and according to the inland or marine area.
Indeed, the data of growth rate for a wide range of production
values plotted in terms of scaled coordinates collapse onto a single
curve (Fig.\ref{fig:4}{\sf B}),
if the ordinate and the abscissa are chosen as
\begin{equation}
 p_{\mbox{\scriptsize sc}}
  \equiv
  \sqrt{2} \sigma_r (x) p(r|x)
  \quad\mbox{and}\quad
  r_{\mbox{\scriptsize sc}}
  \equiv
  \sqrt{2}
  \left[r-\overline{r}(x)\right]/\sigma_r (x),
\end{equation}
respectively.
Accordingly, the probability density function of growth rate $r$ for a
country with $x$ tonnes landings, detrended by the average growth rate
$\overline{r}(x)$,
should scale as
\begin{equation}
 p\left(r'|x\right)\mbox{d}r'
  =
  f_r\left(\frac{r'}{x^{-\beta}}\right)
  \mbox{d}\left(\frac{r'}{x^{-\beta}}\right),
\end{equation}
where the scaling function $f_r$ is exponential, and the detrended growth
rate is denoted by $r'\;[\equiv r-\overline{r}(x)]$.

\subsection{$\Delta t$-year growth rate}
We now question whether Eq.(\ref{eqn:laplace}) of landed catches in
countries is still valid for the growth rates for longer time scales.
Consider the $\Delta t$-year growth rate
\begin{eqnarray}
 r_{\Delta t}
  &\equiv&
  \ln \left[\frac{x(t+\Delta t)}{x(t)}\right] \\
  &=&
   \ln \left[\frac{x(t+1)}{x(t)}\right]
   +\ln \left[\frac{x(t+2)}{x(t+1)}\right]
   +\cdots
   +\ln \left[\frac{x(t+\Delta t)}{x(t+\Delta t -1)}\right] \nonumber.
\label{eqn:T-year-growth-rate}
\end{eqnarray}
Is then $r_{\Delta t}$ for large $\Delta t$ Gaussian?
Figure~\ref{fig:5}{\sf A} shows the distributions of
growth rates for time scales from 1 year up to 30 years
over the entire period (1950--2002) for all countries (excluding China).
In Fig.\ref{fig:5}{\sf B} shown are
the probability distributions of the normalized growth rate defined as
\begin{equation}
 g_{\Delta t}
  \equiv
  \frac{r_{\Delta t}-\overline{r_{\Delta t}}}{\sigma_{\Delta t}},
\end{equation}
where
$\overline{r_{\Delta t}}$ and $\sigma_{\Delta t}$ are the average and
the standard deviation of the $\Delta t$-year growth rate.
We see that
the distributions of $g_{\Delta t}$ are, to a good approximation,
symmetric with respect to zero.
Figure~\ref{fig:5}{\sf C} shows the positive tails of
the complementary cumulative distributions $P(\geq g_{\Delta t})$
of the normalized growth
rates for the same time scales,
compared to the standardized Gaussian distribution with a mean of 0 and
a standard deviation of 1.
Because of the central limit theorem,
the convergence to a Gaussian distribution on longer time scales is
observed,
as the Gibrat model predicts.

We can further test the convergence to Gaussian behavior by analyzing the
moments of the distribution of normalized growth rates.
A more global quantitative measure of distributions is in terms of the
moments:
\begin{equation}
 \mu_k
  \equiv
  \left\langle \left|g_{\Delta t}\right|^k \right\rangle,
\end{equation}
where $\langle\ldots\rangle$ denotes the average.
The fractional moments of the normalized growth rates for
$\Delta t = 1$, 10, 20 and 30 years
(Fig.\ref{fig:5}{\sf D})
show clear indication of convergence to Gaussian behavior with
increasing $\Delta t$.
However, it is remarkable that
they do not satisfyingly approach the Gaussian moments even for large
$\Delta t = 30$ years.

Moreover, notice that the distribution of annual ($\Delta t = 1$)
growth rates $r_1$ for all countries and all years can be well-described
by an exponential (Fig.\ref{fig:5}{\sf A}),
\begin{equation}
 p(r_1)
  =
  \frac 1{\sqrt{2}\sigma_1}
  \exp \left(
	-\frac{\sqrt{2} \left| r-\overline{r_1}\right|}{\sigma_1}
      \right),
  \label{eqn:laplace-entire-period}
\end{equation}
with $\overline{r_1} = 0.04$ and $\sigma_1 = 0.27$,
while the conditional probability density of the yearly growth rates for
countries with approximately the same landings $x$ depends on $x$
[Eqs.(\ref{eqn:laplace}) and (\ref{eqn:scaling-growth-rate}),
Figs.\ref{fig:3} and \ref{fig:4}].
Using a saddle point approximation \citep{Lee98},
we may integrate the conditional Laplace distribution,
Eq.(\ref{eqn:laplace}),
over a log-normal distribution of the country landings,
Eq.(\ref{eqn:lognormal-landings}),
and obtain Eq.(\ref{eqn:laplace-entire-period}).

\section{Scaling structure of the fisheries sector}\label{section:stretched-exponentials}
Let us analyze employment data in the fisheries sector
in order to gain further insight into the organization of fisheries
structure of countries.
Figure~\ref{fig:6}{\sf A},
showing a scatter plot of the number of fishers versus country
landings, suggests that
the conditional probability density $\rho(n|x)$ to find a country with
total catch $x$ landed by $n$ fishers has a long right tail.
I here propose a model for the conditional distributions of fishers
population in terms of ``stretched exponentials''
\citep{Laherrere-Sornette98}:
\begin{equation}
 \rho(n|x)
  =
  \frac{\alpha}{n_0} \left(\frac n{n_0}\right)^{\alpha -1}
  \exp \left[-\left(\frac n{n_0}\right)^{\alpha}\right],
\end{equation}
such shat the complementary cumulative distribution is
\begin{equation}
 P(\geq n|x)
  =
  \int_n^{\infty} \rho (n'|x) \mbox{d}n'
  =
  \exp \left[-\left(\frac n{n_0}\right)^{\alpha}\right],
  \label{eqn:stretched-exponential-CCDF}
\end{equation}
where $n_0$ denotes a reference scale of the distribution,
expected to depend on the landings,
$n_0 = n_0(x)$.
Stretched exponentials are characterized by an exponent $\alpha$ smaller
than one.
The meaning ``reference'' is that from $n_0$ all moments can be
determined, e.g. the means of $n$ and $n^2$ are given by
\begin{equation}
 \overline{n}
  =
  n_0 \alpha^{-1} \Gamma (\alpha^{-1})
  \label{eqn:moment-str-1}
\end{equation}
and
\begin{equation}
 \overline{n^2}
  =
  2n_0^2 \alpha^{-1} \Gamma (2\alpha^{-1}),
  \label{eqn:moment-str-2}
\end{equation}
respectively,
where $\Gamma (q)$ is the gamma function
[equal to $(q-1)!$ for $q$ integer].
When $\alpha$ is small,
$\overline{n}$ will be much larger than $n_0$.
The expression~(\ref{eqn:stretched-exponential-CCDF}) of the stretched
exponential complementary CDF means that
the stretched exponential distribution is qualified by a straight line
when plotting the negative log of complementary CDF,
$-\ln P(\geq n|x)$,
as a function of $n$
on a double-logarithmic scale.
As shown in Fig.\ref{fig:6}{\sf B},
lines appear:
the values of the exponent $\alpha$ range from 0.60 to 0.86,
numerically evaluated from
Eqs.(\ref{eqn:moment-str-1}) and (\ref{eqn:moment-str-2}).
Accordingly,
the fishers population for countries with the same landings
follows a stretched exponential distribution
[aka \citet{Weibull51} distribution with exponent lass than one].

Moreover Fig.\ref{fig:6}{\sf A} suggests that
the typical (i.e. average) number of fishers increases as a power law
with the country landings $x$.
We make the hypothesis that the conditional probability density
$\rho(n|x)$
obey the scaling relation
\begin{equation}
 \rho (n|x) \mbox{d}n
  =
  f_n\left(\frac n{x^{\alpha}}\right) 
  \mbox{d} \left(\frac n{x^{\alpha}}\right),
  \label{eqn:fisher-scaling}
\end{equation}
where $f_n$ denotes a scaling function.
To test the scaling hypothesis in Eq.(\ref{eqn:fisher-scaling}),
we plot the scaled quantities
\begin{equation}
 \rho_{\mbox{\scriptsize sc}}\equiv
  x^{\alpha} \rho (n|x)
  \quad\mbox{vs}\quad
  n_{\mbox{\scriptsize sc}}\equiv
  n/{x^{\alpha}},
  \label{eqn:scaling-hypothesis-fisher}
\end{equation}
and observe the data collapse with a exponent $\alpha = 0.71 \pm 0.01$
(Fig.\ref{fig:7}).
Since the data exhibit scaling, we identify a universal scaling function
fitted with a stretched exponential distribution
\begin{equation}
 f_n(n_{\mbox{\scriptsize sc}})
  =
  \alpha n_{\mbox{\scriptsize sc}}^{\alpha-1}
  \exp \left(-n_{\mbox{\scriptsize sc}}^{\alpha}\right).
  \label{eqn:stretched-exponential}
\end{equation}
Notice that the average number of fishers scales with country
landings $x$ as
\begin{equation}
 \overline{n}(x)
  =
  \int_0^{\infty} n \rho (n|x) \mbox{d}n
  \propto
  x^{\alpha}.
  \label{eqn:typical-number-fishers}
\end{equation}
As shown in Fig.\ref{fig:6}{\sf A} this is indeed the
case.

The multiplicative cascade model \citep{Frisch-Sornette97} may explain
the stretched exponential distribution of fishers:
the probability distribution function of
the product of $m$ independent identically distributed positive random
variables with an exponential-like distribution leads to stretched
exponentials
[see also \citet{Laherrere-Sornette98}].
The reciprocal number of levels in the multiplicative cascade reads the
index $\alpha = m^{-1}$.
Interpreted within the multiplicative cascade model, the value of the
exponent $\alpha\approx 0.7$ corresponds to about one to three generation
levels in the generation of a typical fisher.

Notice that the conditional distribution $\rho (n|x)$ reduces to a
stretched exponential distribution,
while the distribution of the country population engaged in fishing and
fish farming is log-normal (Fig.\ref{fig:2}{\sf B}).
It is numerically demonstrated (Fig.\ref{fig:8}) that
a log-normal distribution of the number of fishers approximately results
from a convolution of a stretched exponential distribution $\rho (n|x)$
and a log-normal distribution of the country landings,
Eq.(\ref{eqn:lognormal-landings}):
\begin{eqnarray}
 P(n)
  & = &
  \int_0^{\infty} \rho (n|x) P(x) \mbox{d}x \nonumber\\
 & = &
  \int_0^{\infty}
  \alpha \frac{n^{\alpha-1}}{n_0^{\alpha}}
  \exp\left[-\left(\frac n{n_0}\right)^{\alpha}\right]
  \ 
  \frac 1{x\sqrt{2\pi s_x^2}}
  \exp
  \left[-\frac{\ln^2 (x/\hat{x})}{2s_x^2} \right]
  \mbox{d}x
\end{eqnarray}
with
$n_0 \propto x^{\alpha}$.

\section{Modeling scaling behavior in the exploitation dynamics}
Total catch of a country is made up of individual fisher activity in the
country.
Fisheries management ranges along a continuum from community to
government as sole managers of the resource, and the field of impact of
their decisions can be local, regional, provincial, or national.
Fishers get involved in their respective management arrangements in
a country.

We now consider the following two limiting cases.
\begin{itemize}
 \setlength{\itemsep}{10pt}
 \item[{(i)}]
There are strong correlations between the fishers in a country,
 following that the growth dynamics are indistinguishable from the
 dynamics of structureless organizations.
As a result, there is no country-size dependence of the growth
 fluctuation $\sigma_r$, following $\beta = 0$.
 \item[{(ii)}]
The annual landed catches by individual fishers change independently of
 one another.
Then, if the number of fishers in a country were proportional to its
 domestic fishery product, the growth fluctuations as a function of
 domestically-landed catch would decay as a power law with an exponent
 $\beta = 0.5$
 \citep{buldyrev-etal97}.
This is because the standard deviation of the sum of $n$ independent
 quantities grows like $\sqrt{n}$ (i.e. catches by $n$ fishers).
\end{itemize}

As to the stock status during the period 1950--1971,
there was much room for continued expansion of capture fisheries
\citep{Grainger-Garcia96}.
The fishing policy those days
would have been virtually that
``{Strive for as much fish as possible from the resources.}''
All the decisions in a country would have been perfectly coordinated
as if they had been all dictated by a single ``boss''.
The state of stock exploitation is then expected to have corresponded to
the case (i):
annual catches by individual fishers in a country were correlated
strongly each other.

In 1972 or subsequent years,
the observed decrease in fluctuations with country landings,
Eq.(\ref{eqn:scaling-growth-rate}),
is considerably slower (i.e. $\beta\approx 0.15$) than in the case~(ii).
Our empirical results suggest an important consequence for dynamics of stock
exploitation:
although countries with large fisheries tend to diversify into a wider
range of fishing activities leading to smaller relative fluctuations in
country landings,
fisheries catches reported are more variable than what would be expected
if the number of fishers would increase linearly with the country
landings---which would correspond to $\beta =0.5$.

\subsection{Scaling relation between exponents}
The country landings does not scale in a simple linear fashion with
increasing number of fishers, but instead
the typical number of fishers scales with landings as
Eq.(\ref{eqn:typical-number-fishers}).
The total capture production is broken down by numbers of landing sites
(i.e. fishing ports) in a country.
The fishery products in landing sites are landed by numbers of local
fishers.
These production equals the marketing volume of fish at local landing
sites.
Let us analyze fishery statistics of Japan on the marketing of fishery
products in major landing areas \citep{MAFF00-02}.
It is virtually apparent in Fig.\ref{fig:9},
showing the probability distribution of catches landed at
203 ports~({\sf A}) and the probability density of annual growth rate of
port landings~({\sf B}),
that the same behavior holds for subsystems, i.e. the single country's
fisheries sector, as the world capture fisheries sector.
The subsystems or hierarchical systems of capture fisheries sector are
expected to be characterized by the same fundamental laws.

To account for the increased variability for highly producing countries
(i.e. $\beta\approx 0.15$ for years 1972--2002),
we suppose that
in a country with $x$ tonnes domestic fisheries production,
the width of the annual per capita growth-rate distribution,
denoted by $\sigma_0(x)$,
scales with landings $x$ as
\begin{equation}
 \sigma_0(x)
  \propto
  x^{\gamma}
  \label{eqn:fluctuation-gamma}
\end{equation}
with an exponent $\gamma$.
Motivated by scaling behavior in the exploitation dynamics,
I expect that
there exist scaling relations among three exponents $\alpha$, $\beta$
and $\gamma$ \citep{Widom65a,Widom65b}.
Assume that the per capita growth rate is independent (uncorrelated),
identically distributed,
though individual fishers may not be right fishing units that produce
catches independently of one another;
e.g. there are $132$ thousand fishery establishments
while $238$ thousand fishers exist in Japan in 2003 \citep{MAFF05}.
The central limit theorem is then applied to give
\begin{equation}
 \sigma_r (x)
  =
  \frac{\sigma_0(x)}{\sqrt{\overline{n}(x)}},
\end{equation}
leading to a scaling relation
\begin{equation}
 \beta
  =
  \frac{\alpha}2 -\gamma.
  \label{eqn:scaling-relationship-beta}
\end{equation}
For $\alpha = 0.71\pm 0.01$ and
$\beta = 0.15\pm 0.02$,
Eq.(\ref{eqn:scaling-relationship-beta}) predicts
$\gamma = 0.21\pm 0.03$.
It is remarkable that,
contrary to country landings (Fig.\ref{fig:4}{\sf A}),
the per capita landings of highly producing countries fluctuate more
than do those of low producing countries
[Eq.(\ref{eqn:fluctuation-gamma}) with $\gamma > 0$]:
fishers in a country with larger landings risk more than do fishers in a
country with smaller landings in capture fisheries.

\section{Discussion}
Studying the biological properties of the FAO global fisheries
statistics on landed catches has given ecological insight to the state
of world fish stocks: e.g.
in Refs \citep{Pauly-etal98,Caddy-Garibaldy00,Pauly-etal05},
global conclusions drawn state that a general decline in mean trophic
level of marine and inland landings has occurred in many regions,
reflecting a change in catch composition from long-lived, high trophic
level, piscivorous bottom fishes toward short-lived, low trophic level
invertebrates and planktivorous pelagic fishes.
In this paper, from the socio-ecological viewpoint,
I have analyzed the FAO data and
found the universal scaling behavior in the dynamics of stock
exploitation.
Moreover, we empirically find
from Fig.\ref{fig:6} and
Eq.(\ref{eqn:typical-number-fishers}),
that
the per capita production of countries with landings $x$ scales as
\begin{equation}
 \frac x{\overline{n}(x)}
  \propto
  x^{1-\alpha},
\end{equation}
showing the increased per capita catches for highly producing countries
(i.e. $1-\alpha > 0$).
However, fishing business may have a high-risk premium:
the larger their expected revenues in highly producing countries,
the larger the fluctuations in annual per capita revenues,
which is predicted by Eq.(\ref{eqn:fluctuation-gamma}) with $\gamma > 0$;
whereas the empirical data demonstrate
decreasing fluctuations in annual domestically-landed catches with
increased country landings
[Fig.\ref{fig:4}{\sf A} and
Eq.(\ref{eqn:scaling-growth-rate}) with $\beta > 0$].
The expected fishing risk (i.e. fluctuations in the individual catches)
in a country with $x$ tonnes annual domestic landings scales with the
individual expected catches:
$\sigma_0(x) \propto [x/\overline{n}(x)]^{\delta}$,
with an exponent $\delta = 0.7\pm 0.1$,
which seems to agree with the value of the exponent $\alpha$.

Though the growth era of the capture fisheries production has been ended
during the last decade,
the world fishers have been growing in number
(35.8 percent from 1990 to 2002).
The fishers population to total world population ratio
has been increasing for the past
thirty years and stood at 60.7 fishers per one thousand people in 2002.
In most developing countries of low and middle-income,
the number of people employed in the fisheries sector has been
growing steadily.
In industrialized economies offering occupational alternatives,
the numbers of fishers have been on a declining trend or at best
stationary.
For instance in two important fishing countries, Japan and Norway,
fishers have decreased in number by 34.3 and 19.7 percent, respectively,
during the 1990 to 2002 period.
The increasing trend of world fishers population
may suggest that
wages received by fishers (mostly in developing countries)
exceed
the level of wages that fishers could expect to earn in alternative
employment opportunities,
as commonly understood in economic analysis from the static viewpoint of
the common-property fishery.
Because fishers would be making money more than their opportunity costs
(i.e. resource rent is positive),
additional fishers would be attracted to the fishery,
until
it reaches the ``bionomic'' equilibrium \citep{Gordon54}
where total sustained revenue equals total sustained cost.
Contrary, if net revenues from fishing are negative, fishers leave the
fishery.

The exploitation dynamics of fish stocks exhibits two-phase behavior,
which may be according to stock status together with management policy.
There was a change in scaling behavior of the standard deviation of
production growth rates at 1972.
It may be more than coincidence that
the emergence of the United Nations' Convention on the Law of the Sea,
in the late 1970s,
which enabled countries to claim exclusive economic zones reaching 200
nautical miles into the open sea,
put the responsibility for fisheries resource management squarely with
maritime countries,
thus ending many decades of fighting over traditional fishing grounds
\citep{Johnstone77}.

Another possible cause for the observed transition at 1972 in scaling
structure is that,
having depleted large valuable stocks,
fishing has redirected some effort and added a lot of it on other
species lower down the food web.
The strategy was advocated in the 1970s to increase fisheries
production
\citep{Sprague-Arnold72,Garcia-Grainger05}.
The phenomenon of change in catch composition, i.e.
``fishing down marine food webs'' \citep{Pauly-etal98},
contrasts strikingly with D'Aancona's observation on Mediterranean
fisheries \citep{D'Ancona26} that
the cessation of the fishery during World War I from 1914--1918 changed
the species composition in favor of fish-eating fishes
(the frequency of predators decreased with an increase in fishing after
the war).
The decline in mean trophic level of the world fish stocks leads to
increasing variability in catches \citep{Pauly-etal02}.
Indeed, we find from the analysis of the FAO data for all countries
(excluding China) that
the country landings fluctuate more for the periods 1972--2002
than do those for the periods 1950--1971.
The standard deviation of growth rate of the annual domestically-landed
catches for 30 one-year periods between 1972 and 2002
shows an increase of more than 10 percent over that for 21 one-year
periods between 1950 and 1971:
\begin{equation}
\overline{\sigma}_{r,\mbox{\tiny 1950--1971}} = 0.138\pm 0.004,\quad
\overline{\sigma}_{r,\mbox{\tiny 1972--2002}} = 0.156\pm 0.003,
\end{equation}
whereas the average yearly growth rate declined significantly after
1972:
\begin{equation}
\overline{r}_{\mbox{\tiny 1950--1971}} = 0.077,\quad
\overline{r}_{\mbox{\tiny 1972--2002}} = 0.026.
\end{equation}
Besides highly variable recruitment of planktivorous pelagic fishes that
is strongly influenced by environmental fluctuations,
the schooling behavior common among such pelagics is a possible cause of
variation in catch in purse seine fisheries.
Some fishers catch more than others:
the distribution of fishers' yearly total landed value of catch in a
commercial purse seine fishery is skewed to the right \citep{Hilborn85}.
School-size distributions of pelagic fishes, i.e. catch-per-haul
distributions in purse seine fisheries, are generally heavy-tailed,
and a typical school-size
(rigorously defined as a crossover size from power-law to exponential
decay of the school-size distribution)
is proportional to the abundance of the fish
stock \citep{Niwa03,Niwa04}, seeming to be consistent with
Eq.(\ref{eqn:fluctuation-gamma}) with $\gamma > 0$.

The global potential harvest (e.g. maximum sustainable yield) may be one
of the causes for the observed transition in scaling behavior of fishery
landings.
\citet{Grainger-Garcia96} estimated the most likely potential of
conventional marine species to range from 80 to 100 million tonnes,
which has indeed been reached probably in the 1970s and is unlikely to
change in the next 20--30 years \citep{Garcia-Grainger05}.
Undeveloped resource fisheries, producing much less than their
potential, decreased rapidly to zero by the middle of the 1970s
\citep{Grainger-Garcia96}.
The latest analysis of the state of resources \citep{Garcia-etal05}
indicates that, in 2003, approximately half the world's stocks are
exploited at or close to their maximum, and about 25 percent of them are
exploited either below or above such maximum.

The same scaling laws as found here describe a number of other complex
organizations \citep{Stanley02}, such as
business firms \citep{Stanley96},
gross domestic product \citep{Lee98},
university research budgets \citep{Plerou99},
and
bird populations \citep{Keitt-etal02},
suggesting the underlying universal mechanism.
It is no wonder that commercial fisheries world-wide resemble
competitive economic activities.
In the same way that market forces operate in business, 
the stock exploitation appears to keep competition among fishers
strong enough to leave capture fisheries with a shrinking resource base.
According to FAO press release \citep{double},
most small-scale fishers find it increasingly difficult to survive in an
over-exploited environment.

\citet{Amaral98} proposed a model for the growth of organizations with
complex internal structure.
In the model, each organization
(e.g. firm, university, or bird population)
is made up of units.
The units composing the system have a complex evolving structure,
e.g.
business firms competing in an economy are comprised of divisions;
universities are composed of schools or colleges;
population of breeding birds living in a given ecosystem is composed
of groups living in different areas.
Their model postulates that
the units grow through an independent, Gaussian-distributed, random
multiplicative process.
Later this postulate empirically proved not to be accurate
\citep{Matia-etal04}.
Thus the model proposed by \citet{Amaral98} can now be regarded
as a first step towards the explanation of the growth dynamics of
organizations.
Scaling concepts offer an avenue to study heterogeneous assemblies in
ecological and economic systems for which the microscopic processes are
(probably) not knowable, except in terms of their statistical properties
\citep{Chave-Levin03}.

\newpage

%

\newpage
\begin{flushleft}
{\bf Figure legends}
\end{flushleft}

\begin{itemize}
 \setlength{\itemsep}{10pt}
\item[Fig.1]
 World fisheries trend including ($\circ$)/excluding ($\bullet$) China.
 ({\sf A})~Semi-logarithmic scale plot of global capture trend.
 Exponentially growing behavior with two growth-rates is observed.
 There is a knee in the curve at 1972,
 probably figuring the boundary between two distinct regimes of the
 bio-economies.
 The solid lines are fits of an exponential function
[$\propto \exp (Rt)$]
 to the data~($\bullet$) with average yearly growth rates
 $R = 0.059\pm 0.001$ and $0.024\pm 0.001$
 for the periods 1950--1971 and 1972--1989, respectively.
 A plateau in the 1990s follows immediately
 (to be precise $R = -0.002\pm 0.003$).
 Fitting the data~($\circ$) gives
 $R = 0.058\pm 0.001$ and $0.017\pm 0.001$
 for the periods 1950--1971 and 1972--2002, respectively
 (depicted by broken lines).
 ({\sf B})~Trend of the world fisher population $N$.
 The ordinate represents $\log_{10} N$.
 The broken and solid lines depict exponential fits to the data
 including and excluding China,
 giving population growth rate
 $R_N = 0.035\pm 0.002$ and
 $R_N = 0.028\pm 0.001$
 for the period 1970--2002, respectively
 (annual growth rate of the fisher population world-wide is defined as
 same as the global harvest case).
 The inserted figure shows the fisher-to-population ratio.
 The earth's human population data from 1970 to 2000 were retrieved from
 the web-site of \citet{unpp04}.
 The human population has been growing at an average rate
 of 1.67 percent per year (excluding China, 1.74 percent)
 since 1970 to 2000.
\item[Fig.2]
 Fisheries production and employment statistics.
 Data exclude China's statistics.
 ({\sf A})~Semi-logarithmic scale plot of probability distribution of
 country landings,
 $P(x)$.
 The abscissa is chosen as logarithmically scaled landings
 $\ln (x/\hat{x})/s_x$.
 The data from 6 one-year periods between 1950 and 2000 are plotted,
 where geometric means $\hat{x}$
 of landings (in thousand tonnes) and geometric standard deviations $s_x$
 are as follows:
 $\hat{x} = 6.0$,
 9.8,
 15.4,
 16.9,
 20.1 and
 20.4, and
 $s_x = 2.5$,
 2.6,
 2.7,
 2.9,
 2.9 and
 2.9,
 from years
 1950~$(\square)$,
 1960~$(\lozenge)$,
 1970~$(\vartriangle)$,
 1980~$(\blacksquare)$,
 1990~$(\blacklozenge)$ and
 2000~$(\blacktriangle)$, respectively.
 Log-scaled landings fall onto
 the standardized Gaussian distribution (dashed line)
 with zero mean and unit variance.
 ({\sf B})~Semi-logarithmic scale plot of probability distributions of
 the number of fishers and fish farmers in countries,
 $P(n)$.
 The abscissa is chosen as logarithmically scaled fishers population
 $\ln (n/\hat{n})/s_n$,
 where geometric means $\hat{n}$
 of the number of fishers (in thousand fishers) and geometric standard
 deviations $s_n$ are as follows:
 $\hat{n} = 6.7$,
 8.1 and
 10.2, and
 $s_n = 2.1$,
 2.1 and
 2.0,
 from years
 1970~$(\vartriangle)$,
 1980~$(\blacksquare)$ and
 1990~$(\blacklozenge)$, respectively.
 Log-scaled fishers population fall onto
 the standardized Gaussian distribution (dashed line).
 The log-normal distribution holds robustly for fishers as well as landings.
\item[Fig.3]
 Exploitation dynamics of fish stocks.
({\sf A})~Semi-logarithmic scale plot of probability density $p(r|x)$ of
 the annual growth rate for eight different bins of
 fisheries production (in tonnes):
 $x < 10^{2.5}\; (\circ)$,
 $10^{2.5} \leq x <10^3\; (\square)$,
 $10^3 \leq x <10^{3.5}\; (\lozenge)$,
 $10^{3.5} \leq x < 10^4\; (\vartriangle)$,
 $10^4 \leq x <10^{4.5}\; (\blacksquare)$,
 $10^{4.5} \leq x < 10^5\; (\blacklozenge)$,
 $10^5 \leq x < 10^{5.5}\; (\blacktriangle)$,
 $10^{5.5} \leq x\; (\bullet)$.
 The data are aggregated over all 21 one-year periods between 1950 and
 1971.
 The distribution decays with fatter tails than for a Gaussian.
 The broken line is a fit of a Laplace distribution
 [Eq.(\ref{eqn:laplace})] for the period, with
 $\overline{r} =0.077$ and $\sigma_r = 0.138\pm 0.004$.
 It is depicted that $p(r|x)$ does not depend on $x$.
({\sf B})~Semi-logarithmic scale plot of probability density of the
 growth rate for countries with small and large landings
 in the years 1972--1989/1990--2002:
 $10^3 \leq x <10^4\; (\lozenge /\vartriangle)$ and
 $10^5 \leq x < 10^6\; (\blacklozenge /\blacktriangle)$ in tonnes.
 The broken and solid lines are Laplacian fits using the means
 $\overline{r} = 0.035$ and $0.016$
 and standard deviations
 $\sigma_r = 0.203\pm 0.004$ and $0.117\pm 0.004$,
 calculated from data for countries with small and large landings in the
 years 1972--1989, respectively.
 Fits with Eq.(\ref{eqn:laplace})
 for small and large landings in 1990--2002
 give the means
 $\overline{r} = 0.034$ and $0.005$
 and standard deviations
 $\sigma_r = 0.219\pm 0.009$ and $0.118\pm 0.007$, 
 overlapping broken and solid lines in the graph.
\item[Fig.4]
 Scaling in stock exploitation dynamics.
({\sf A})~Standard deviation of the one-year growth rates of the 21 annual
 intervals from the period 1950--1971~($\circ$), of the 17 annual
 intervals from the period 1972--1989~($\bullet$), and of the 12 annual
 intervals from the period 1990--2002~($\blacksquare$), as a function of
 the landings $x$
 (plotted on a double-logarithmic scale).
 Each data point indicates the width $\sigma_r (x)$ of the growth-rate
 distribution for countries with approximately the same landings
 (binned evenly spaced on a logarithmic scale).
 The long-dashed horizontal line depicts the average of
 the growth fluctuations for years 1950--1971 ($\sigma_r = 0.138$),
 showing $\beta = 0$.
 The solid and broken lines are fits with
 Eq.(\ref{eqn:scaling-growth-rate}) to the data from
 years 1972--1989 ($\beta = 0.14\pm 0.02$) and
 1990--2002 ($\beta = 0.15\pm 0.02$), respectively.
 There are cutoffs at the small landings in power laws.
 Error bars of one standard deviation about each data point are shown.
({\sf B})~Scaled probability density
 $p_{\mbox{\scriptsize sc}}$
 as a function of the scaled growth rate
 $r_{\mbox{\scriptsize sc}}$
 for seven different bins of production in the years 1972--1989 (in
 tonnes):
 $10^3 \leq x <10^{3.5}\; (\square)$,
 $10^{3.5} \leq x < 10^4\; (\lozenge)$,
 $10^4 \leq x <10^{4.5}\; (\vartriangle)$,
 $10^{4.5} \leq x < 10^5\; (\blacksquare)$,
 $10^5 \leq x < 10^{5.5}\; (\blacklozenge)$,
 $10^{5.5} \leq x < 10^6\; (\blacktriangle)$,
 $10^6 \leq x\; (\bullet)$.
 The values are scaled by using the measured values of
 $\overline{r}(x)$ and $\sigma_r (x)$,
 and plotted on a semi-logarithmic scale.
 All the data collapse upon the universal curve
 $p_{\mbox{\scriptsize sc}} = \exp
 \left(-\left|r_{\mbox{\scriptsize sc}}\right|\right)$
 (broken line).
 A data collapse can be obtained for 1990--2002 as well.
\item[Fig.5]
 Growth rates of country landings for longer time scales.
 The data points are aggregated over the entire period, 1950--2002,
 for all countries (excluding China).
 ({\sf A})~Probability densities $p(r_{\Delta t})$ of $\Delta t$-year
 growth rates are shown
 for $\Delta t = 1$ ($\square$),
 10 ($\lozenge$),
 20 ($\vartriangle$)
 and
 30 years ($\blacksquare$),
 plotted on a semi-logarithmic scale.
 Averages and standard deviations of the $\Delta t$-year growth rate
 are as follows:
 $\overline{r_{\Delta t}} = 0.04$, 0.39, 0.73 and 1.08,
 and 
 $\sigma_{\Delta t} = 0.27$, 0.80, 0.99 and 1.11
 for the same time scales.
 The solid line indicates a Laplace distribution
 [Eq.(\ref{eqn:laplace-entire-period})] with
 $\overline{r_1}$ and $\sigma_1$.
 It is visually apparent that the distributions approach a Gaussian as
 $\Delta t$ increases.
 ({\sf B})~Semi-logarithmic scale plot of probability distributions of
 the normalized growth rates $g_{\Delta t}$.
 The broken line shows the standardized Gaussian distribution.
 ({\sf C})~Double-logarithmic scale plot of complementary cumulative
 distributions
 $P(\geq g_{\Delta t})$ for the positive tails
 of the probability densities shown in Fig.{\sf B}.
 The complementary CDFs for the negative tails have the same functional
 forms as for the positive tails.
 ({\sf D})~Fractional moments for the normalized growth rates for the
 same time scales:
 $\Delta t = 1$ (long-dashed line),
 10 (medium-dashed line),
 20 (short-dashed line)
 and
 30 years (dotted line).
 The solid line shows the Gaussian moments.
\item[Fig.6]
 Statistical analysis of the number of fishers.
 Data exclude China's statistics.
 ({\sf A})~Number of fishers versus country landings (in tonnes).
 The ordinate and the abscissa are chosen as
 $\log_{10} n$ and $\log_{10} x$, respectively.
 Data are aggregated for seven years (1990--1996).
 The rectangle in gray reads the inter-quantile range of binned data
 and the slit indicates the median,
 showing that the conditional distribution $\rho (n|x)$ is skewed to the
 right,
 where the bins are chosen equally spaced on a logarithmic scale as
 $x\in [10^{k-0.5},10^{k+0.5})$ with $k=3,\ldots,6$.
 The dashed line shows a power-law fit to scatter plot data,
 suggesting that the typical number of fishers scales with country
 landings as $x^{\alpha}$.
 From the slope of the dashed line, the exponent $\alpha$ is evaluated at
 $0.68 \pm 0.08$.
 ({\sf B})~Stretched exponential law in the conditional probability of
 $n$ given $x$, $\rho (n|x)$.
 The negative logs of complementary CDF,
 $-\ln P(\geq n|x)$,
 are plotted versus decimal log of the number of fishers,
 $\log_{10} n$,
 on a semi-logarithmic scale.
 The data points fall onto straight lines.
 The countries are partitioned into four groups
 according to landings (in tonnes):
 $10^{2.5} \leq x < 10^{3.5}\; (\lozenge)$,
 $10^{3.5} \leq x < 10^{4.5}\; (\vartriangle)$,
 $10^{4.5} \leq x < 10^{5.5}\; (\blacklozenge)$,
 $10^{5.5} \leq x < 10^{6.5}\; (\blacktriangle)$.
 The solid lines are stretched exponential fits to the data,
 representing power-law curves,
 $-\ln P(\geq n|x) = (n/{n_0})^{\alpha}$,
 where
 $\alpha = 0.86\; (\lozenge)$,
 $0.72\; (\vartriangle)$,
 $0.66\; (\blacklozenge)$ and
 $0.60\; (\blacktriangle)$,
 estimated from
 Eqs.(\ref{eqn:moment-str-1}) and (\ref{eqn:moment-str-2}).
 The estimated reference scales $n_0$ are depicted in Fig.{\sf A} with
 symbol ($\circ$), which closely follow the dashed line.
 The exponent characterizing stretched exponential distributions should
 be the exponent $\alpha$ in the power-low dependence of the average
 number $\overline{n}(x)$ of fishers in a country with annual domestic
 landings $x$ as suggested in Fig.{\sf A}
 (dashed line).
 A power-law fit with $n_0 (x) = B x^{\alpha}$
 to estimated $n_0$'s (solid line in Fig.{\sf A}) gives
 $\alpha = 0.71 \pm 0.01$ and $B = 11 \pm 2$,
 exhibiting a good agreement with values of the exponent estimated from
 Eqs.(\ref{eqn:moment-str-1}) and (\ref{eqn:moment-str-2}).
\item[Fig.7]
 Scaling in employment statistics for the period 1990--1996.
 ({\sf A})~To test the scaling hypothesis in
 Eq.(\ref{eqn:fisher-scaling}),
 the scaled probability distributions
 $\rho_{\mbox{\scriptsize sc}}\;(=B x^{\alpha} \rho)$
 are plotted
 versus scaled number
 $n_{\mbox{\scriptsize sc}}\;(=n/Bx^{\alpha})$ of fishers
 (double-logarithmic scale).
 The data for four different bins are scaled by using the values of
 landings
 $x=10^3\; (\lozenge)$,
 $10^4\; (\vartriangle)$,
 $10^5\; (\blacklozenge)$,
 $10^6$ tonnes ($\blacktriangle$),
 with $\alpha = 0.71$ and $B = 11$.
 The conditional PDF $\rho (n|x)$ exhibits scaling, i.e.
 the data collapse onto a single curve,
 yielding a scaling function $f_n$.
 The solid line depicts the stretched exponential distribution
 [Eq.(\ref{eqn:stretched-exponential})].
 ({\sf B})~Shown is the data collapse of the four curves
 in Fig.\ref{fig:6}{\sf B}.
 The negative logs of complementary CDF,
 $-\ln P(\geq n_{\mbox{\scriptsize sc}})$,
 are plotted
 versus scaled number $n_{\mbox{\scriptsize sc}}$ of fishers
 (double-logarithmic scale).
 The solid line depicts the power-law curve,
 $-\ln P(\ge n_{\mbox{\scriptsize sc}}) =
 n_{\mbox{\scriptsize sc}}^{\alpha}$.
\item[Fig.8]
 Recovering a log-normal distribution from a stretched exponential.
 Semi-logarithmic scale plot of
 probability distribution of the number of fishers for all countries
 (excluding China).
 The abscissa is chosen as $\log_{10} n$.
 The data points are the average over the period 1990--1996, and
 the dashed line is a fit of the log-normal PDF to the
 data with geometric mean
 $\hat{n} = 12.9$ (in thousand fishers)
 and geometric standard deviation
 $s_n = 2.09$.
 Numerically integrating the conditional PDF $\rho (n|x)$ using
 stretched exponential function Eq.(\ref{eqn:stretched-exponential})
 with $\alpha = 0.71$
 over a PDF of the country landings using
 Eq.(\ref{eqn:lognormal-landings}),
 a log-normal distribution of the number of fishers is approximately
 recovered (solid line).
 The reference size of the stretched exponential distribution
 $\rho (n|x)$ is $n_0 (x) = Bx^{\alpha}$ with $B = 11$.
 The distribution of the landing data for years 1990--1996 has
 the geometric mean
 $\hat{x} = 17.1$ (in thousand tonnes) and
 the geometric standard deviation
 $s_x = 3.22$.
\item[Fig.9]
 Statistics on the marketing of fishery products in landing
 sites (203 ports) in Japan.
 ({\sf A})~Semi-logarithmic scale plot of probability distribution of
 local capture production $\xi$.
 The abscissa is chosen as logarithmically scaled landings
 $\ln (\xi/\hat{\xi})/{s_{\xi}}$,
 where geometric means $\hat{\xi}$
 of landings (in thousand tonnes) and geometric standard deviations
 $s_{\xi}$ are as follows:
 $\hat{\xi} = 4.01$,
 3.60 and
 3.33, and
 $s_{\xi} = 1.75$,
 1.70 and
 1.66,
 from years
 2000 $(\square)$,
 2001 $(\blacksquare)$ and
 2002 $(\lozenge)$, respectively.
 Log-scaled landings fall onto
 the standardized Gaussian distribution (dashed line).
 The port landing data exhibit the log-normal probability distribution,
 same as the distribution of country landings (Fig.\ref{fig:2}{\sf A}).
 ({\sf B})~Semi-logarithmic scale plot of probability density function
 of annual growth rate $r_{\xi}$ of port landings.
 The points are the average over the years 2000--2002.
 We find the same tent shape distribution as
 the distribution of country landings growth rate plotted on
 a semi-logarithmic scale.
 The solid line is an exponential fit,
 $
 p(r_{\xi}) =
 (\sqrt{2}\sigma_{\xi})^{-1}
 \exp \left(-\sqrt{2} |r_{\xi}-\overline{r}_{\xi}|/\sigma_{\xi}\right)
 $,
 with average growth rate $\overline{r}_{\xi} = -0.06$
 and standard deviation $\sigma_{\xi} = 0.30\pm0.02$.
 The dashed line is a Gaussian fit with standard deviation
 $\sigma_{\mbox{\tiny Gauss}} = 0.49$.
\end{itemize}

\newpage
\begin{flushleft}
{\bf Figures}
\end{flushleft}

\begin{itemize}
 \setlength{\itemsep}{5pt}
 \item[Fig.1]
 World fisheries trend including ($\circ$)/excluding ($\bullet$) China.
 \item[Fig.2]
 Fisheries production and employment statistics.
 \item[Fig.3]
 Exploitation dynamics of fish stocks.
 \item[Fig.4]
 Scaling in stock exploitation dynamics.
 \item[Fig.5]
 Growth rates of country landings for longer time scales.
 \item[Fig.6]
 Statistical analysis of the number of fishers.
 \item[Fig.7]
 Scaling in employment statistics for the period 1990--1996.
 \item[Fig.8]
 Recovering a log-normal distribution from a stretched exponential.
 \item[Fig.9]
 Statistics on the marketing of fishery products in landing
 sites in Japan.
\end{itemize}

\newpage
%
%

\begin{figure}[p]
 \centering
 \includegraphics[width=6.5cm]{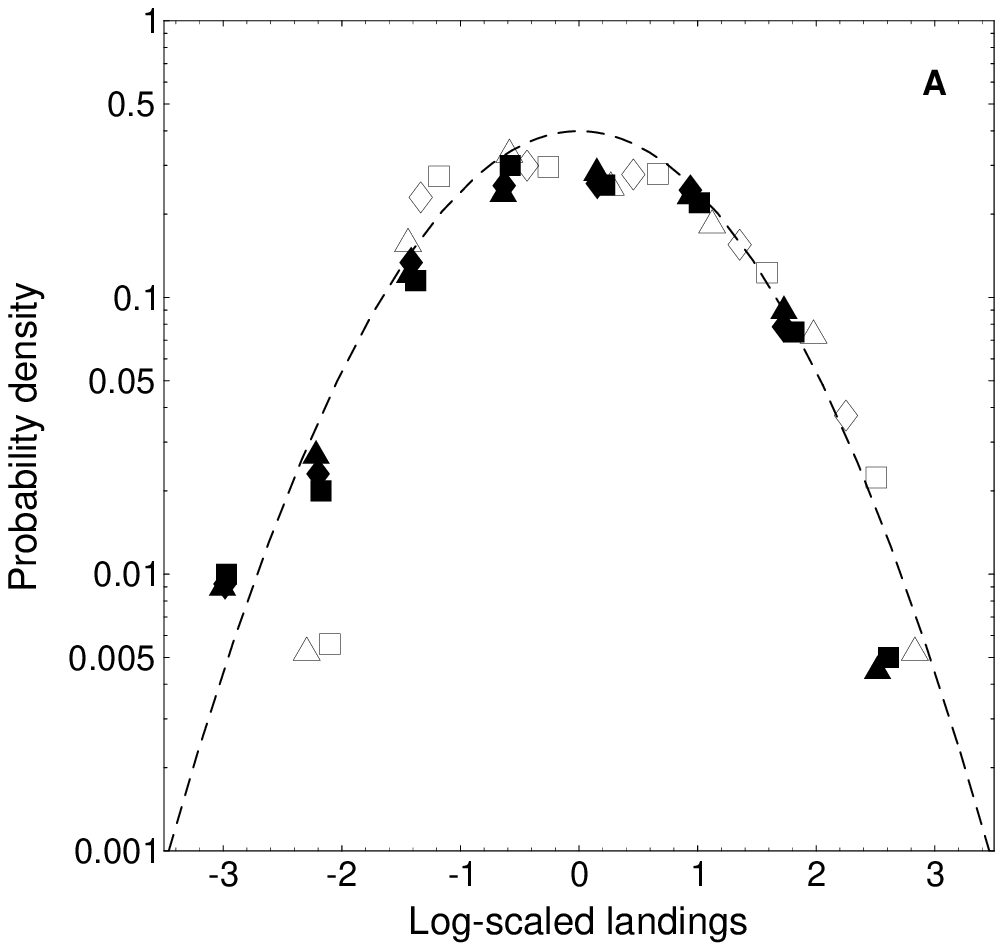}
 \hspace{0.5cm}
 \includegraphics[width=6.5cm]{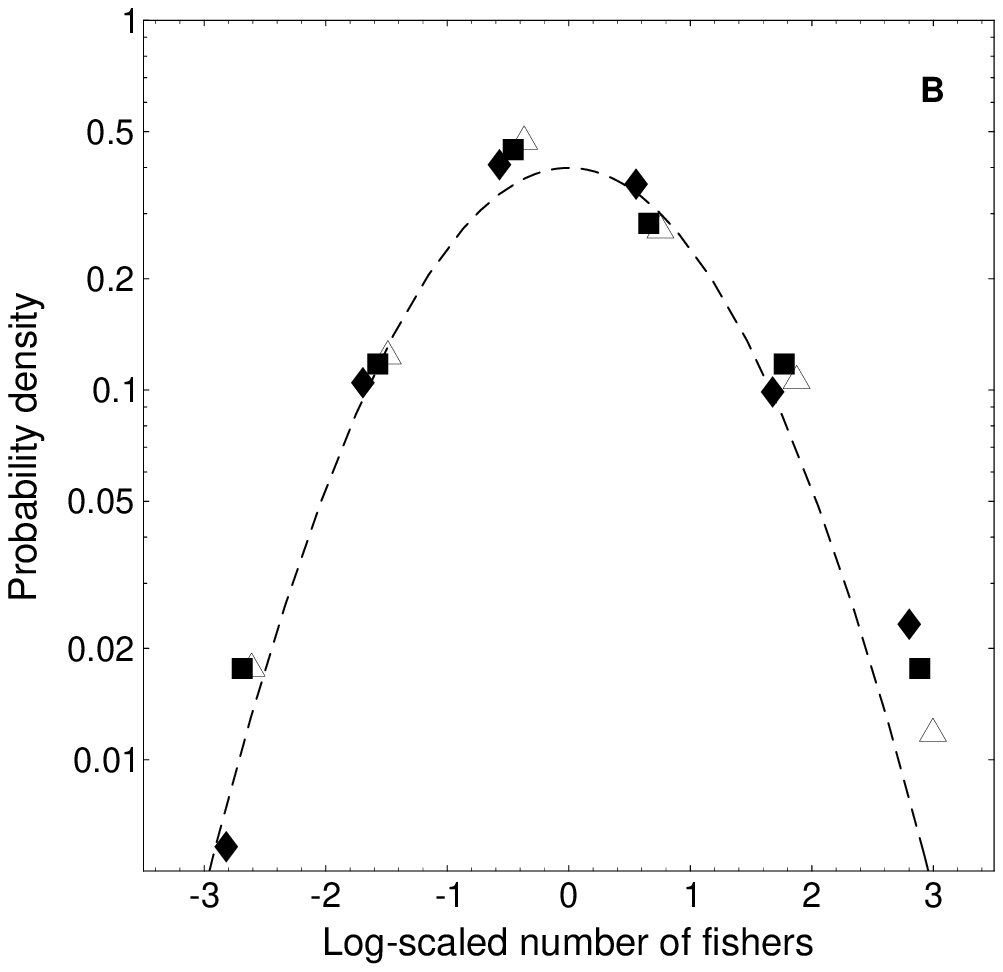}
 \caption{}
 \label{fig:2}
\end{figure}
%

\begin{figure}[p]
 \centering
{
 \includegraphics[width=6.5cm]{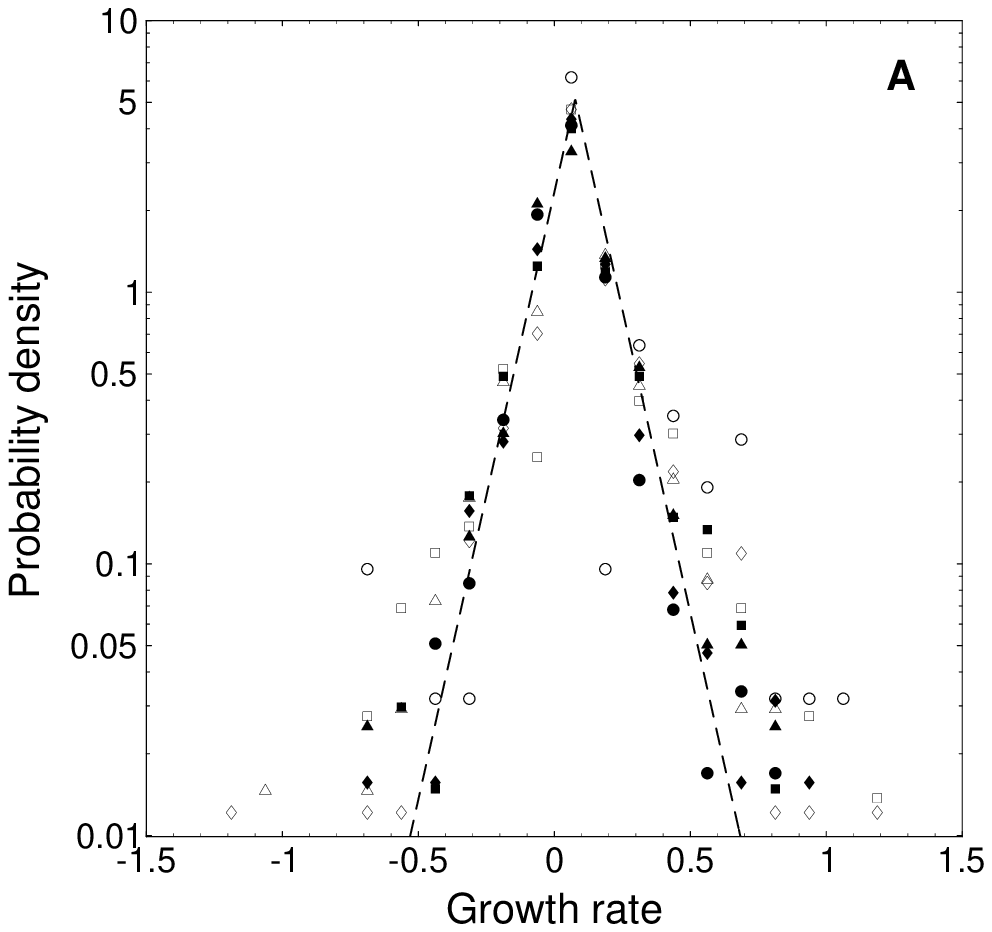}
 \hspace{0.5cm}
 \includegraphics[width=6.5cm]{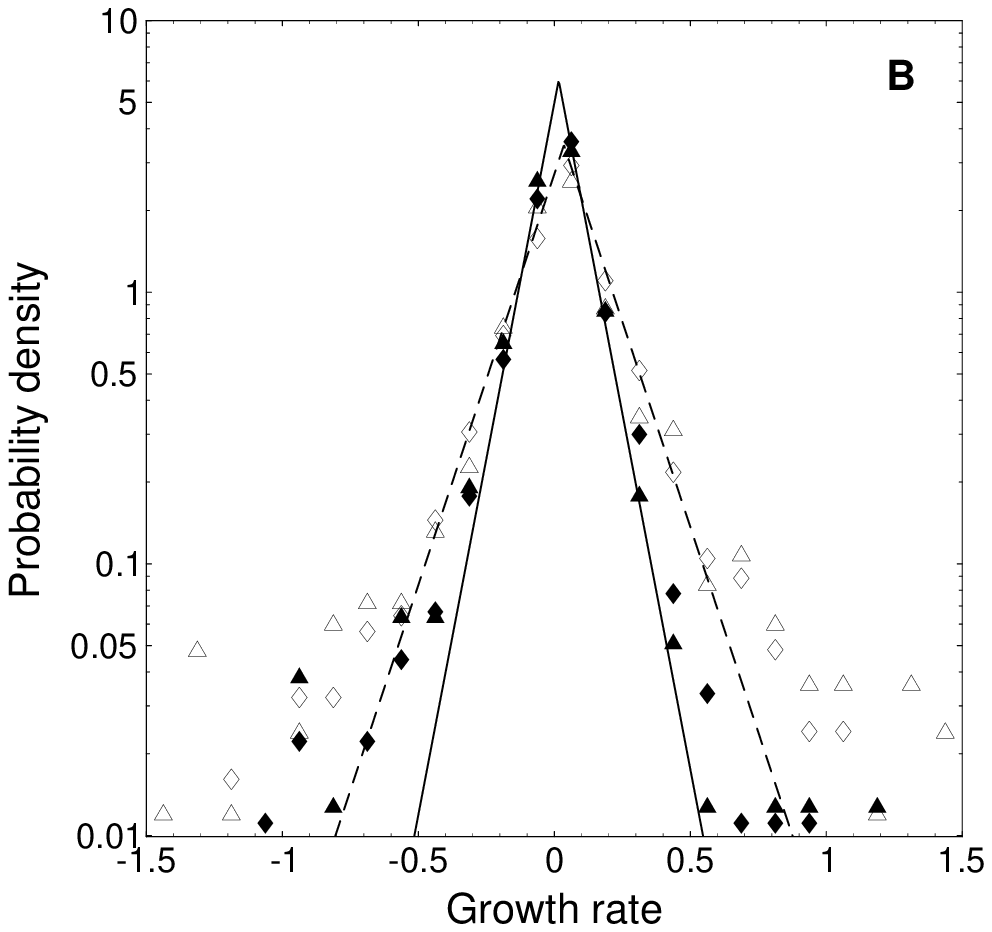}
}
 \caption{}
 \label{fig:3}
\end{figure}
%

\begin{figure}[p]
 \centering
{
 \includegraphics[width=6.5cm]{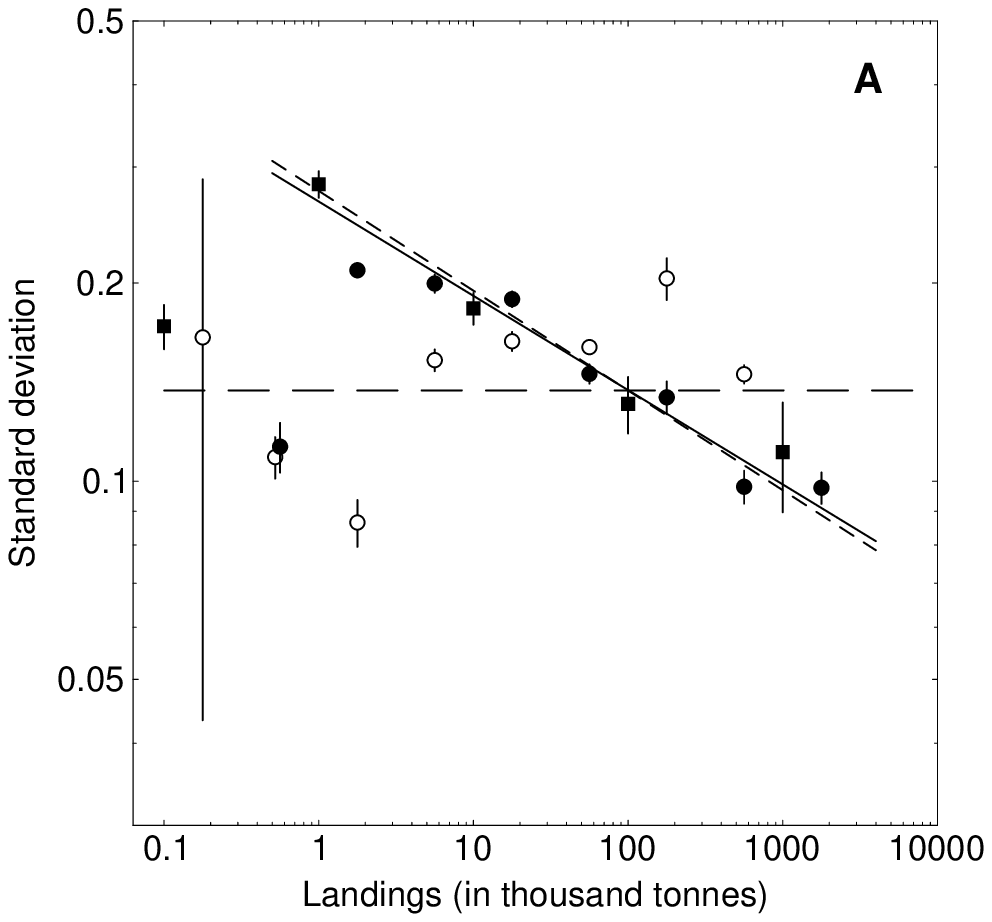}
 \hspace{0.5cm}
 \includegraphics[width=6.5cm]{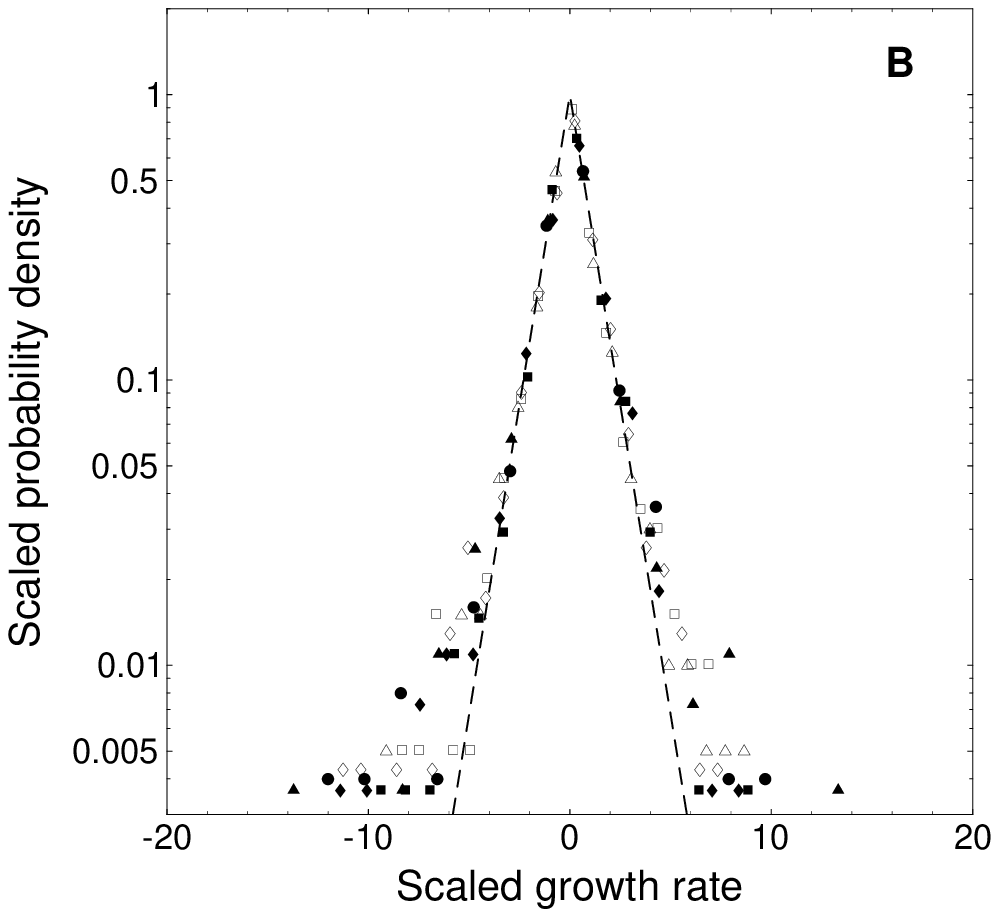}
}
 \caption{}
 \label{fig:4}
\end{figure}

\begin{figure}[p]
 \centering
{
 \includegraphics[width=6.5cm]{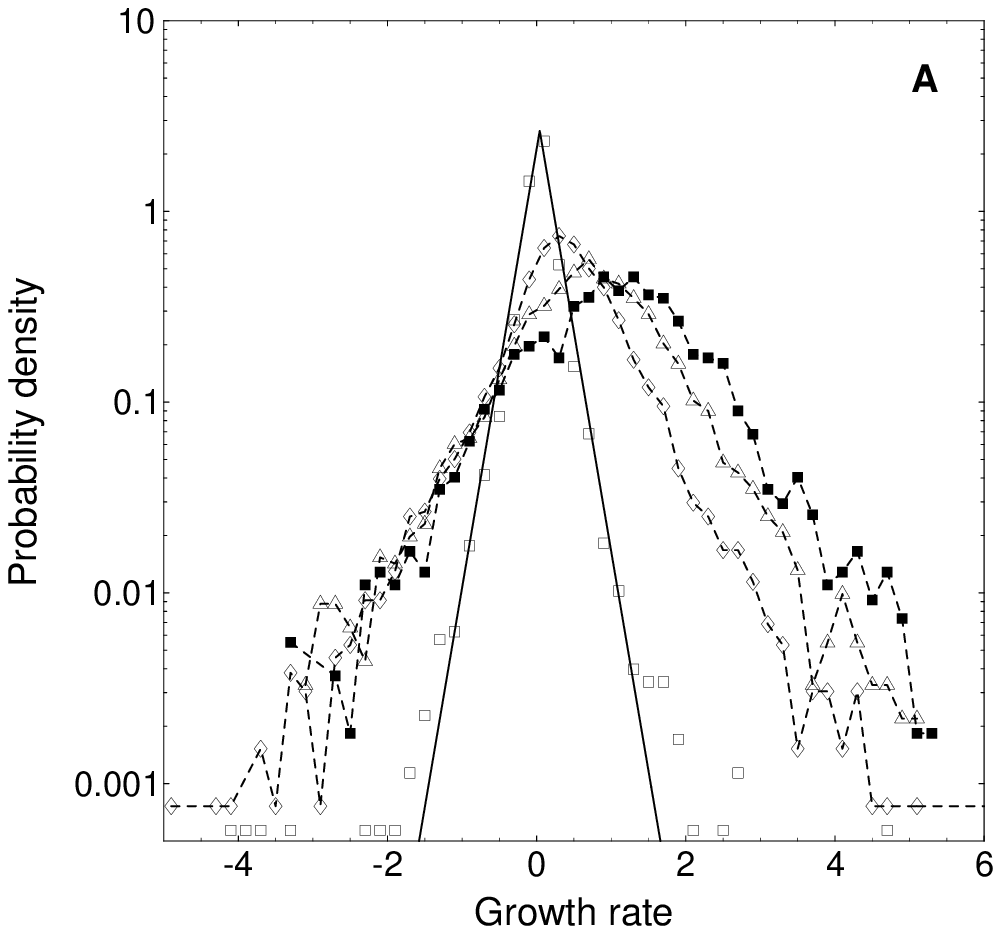}
 \hspace{0.5cm}
 \includegraphics[width=6.5cm]{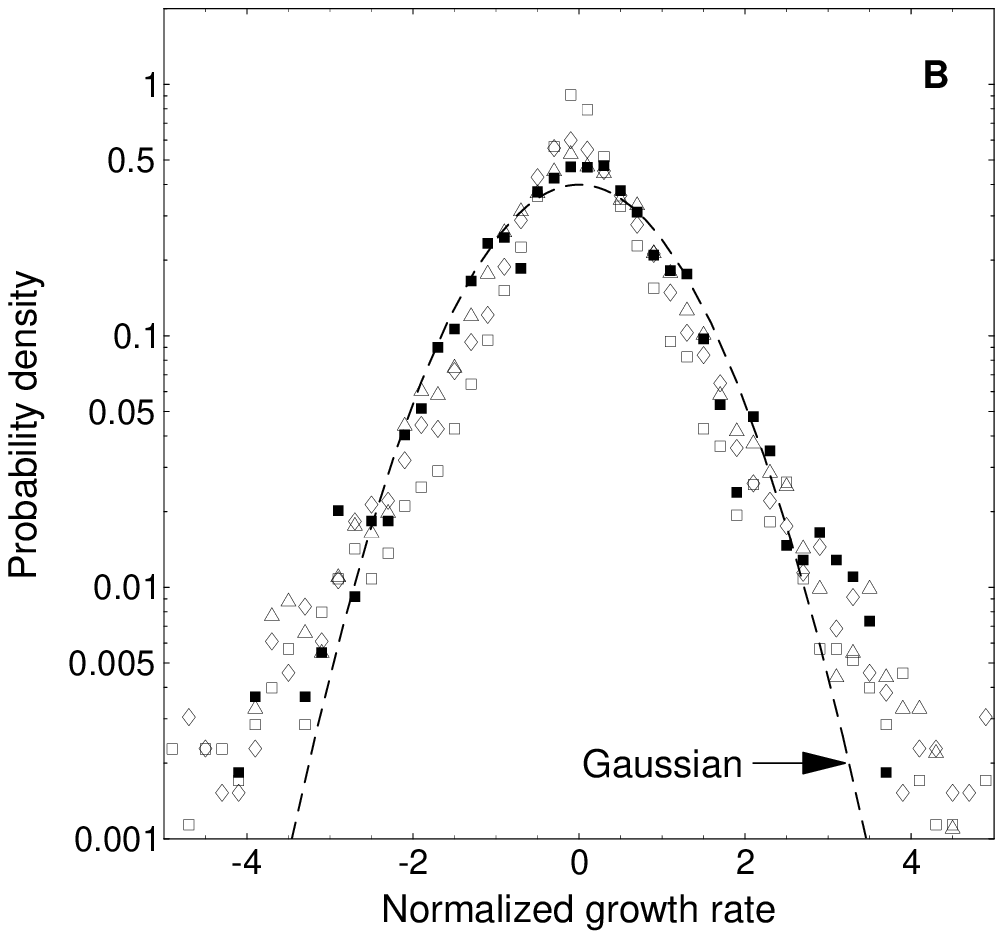}
}
\vspace{0.0cm}
{
 \includegraphics[width=6.5cm]{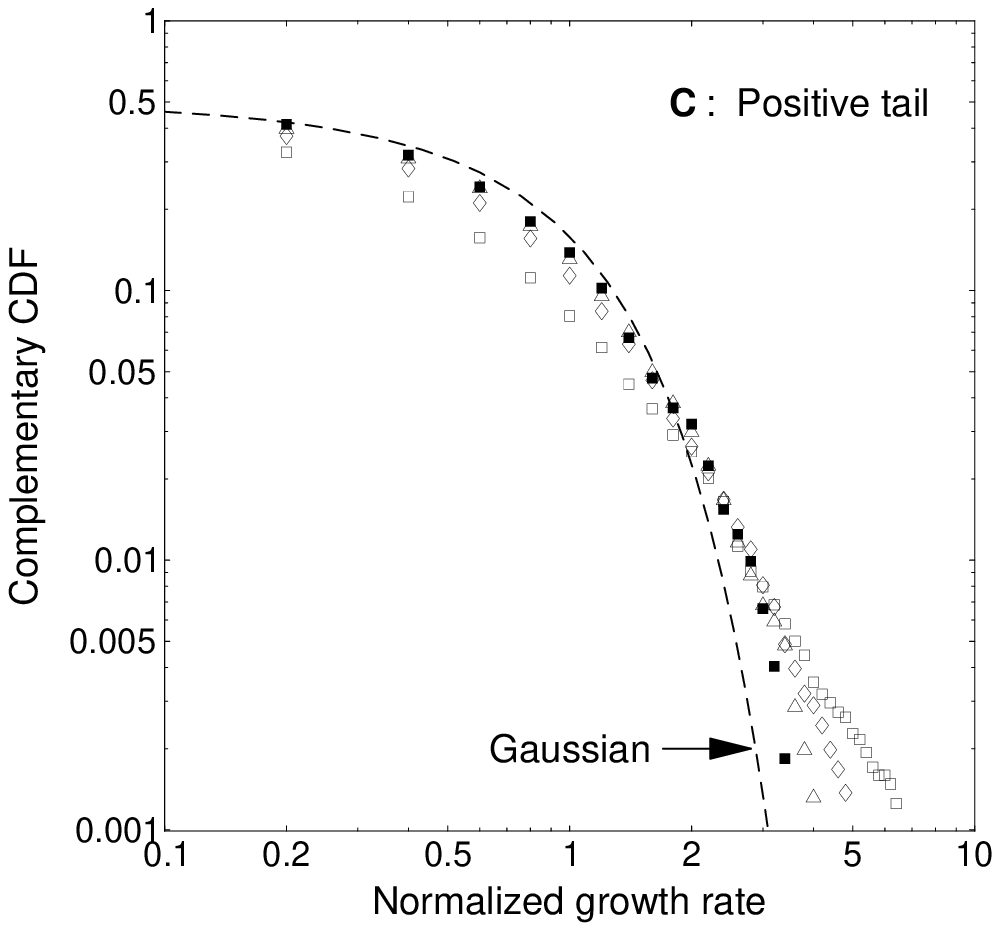}
 \hspace{0.5cm}
 \includegraphics[width=6.5cm]{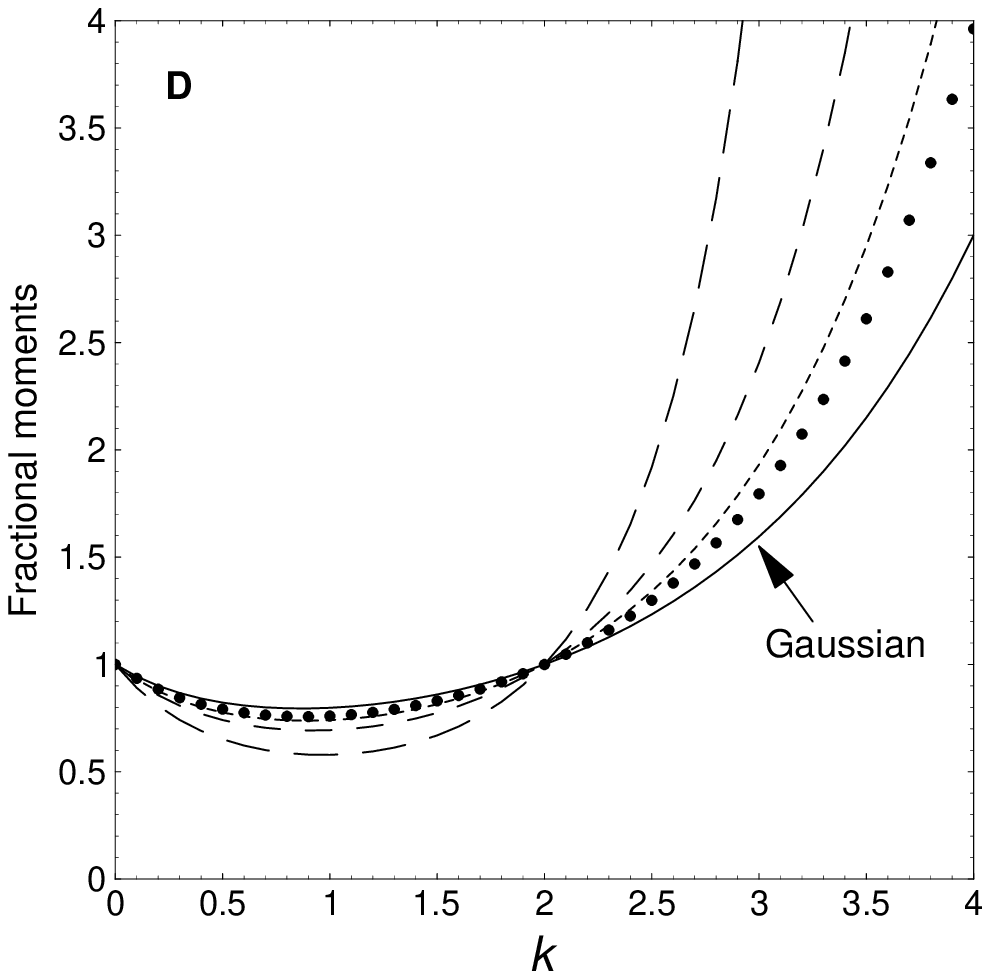}
}
 \caption{}
 \label{fig:5}
\end{figure}
%

\begin{figure}[p]
 \centering
{
 \includegraphics[width=6.5cm]{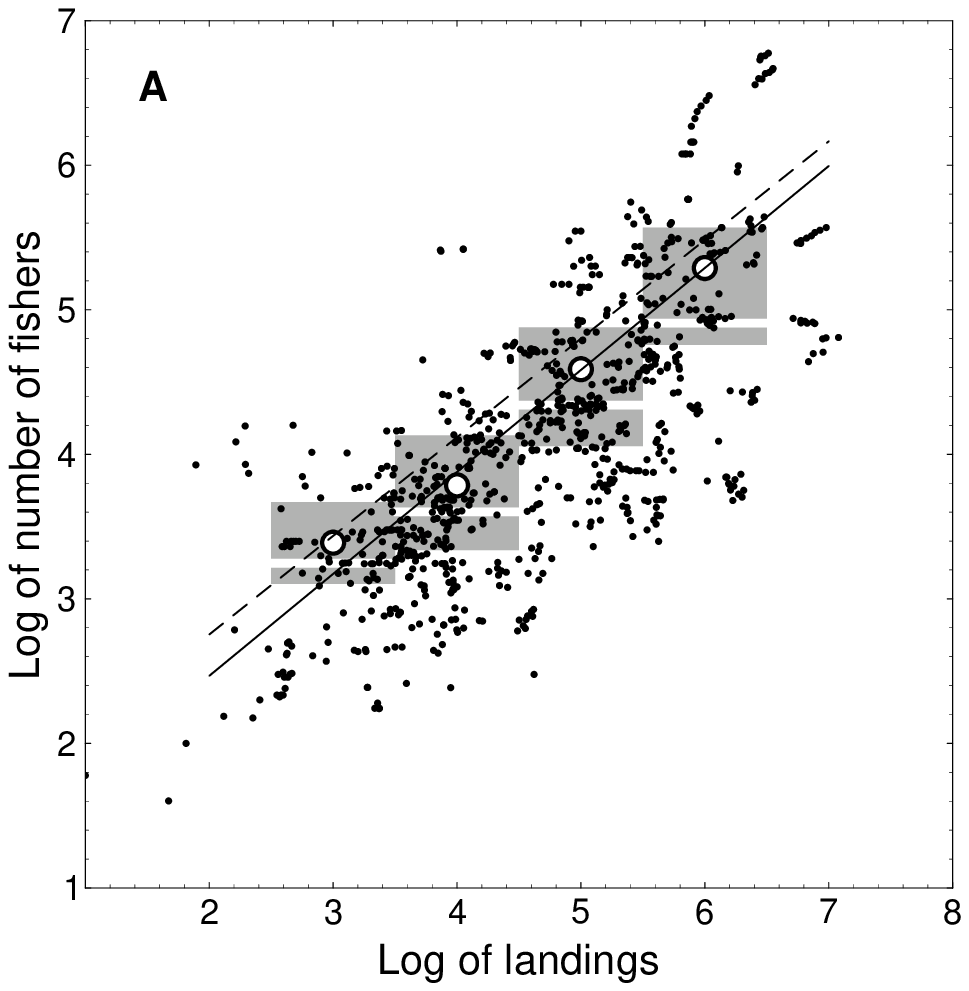}
 \hspace{0.5cm}
 \includegraphics[width=6.5cm]{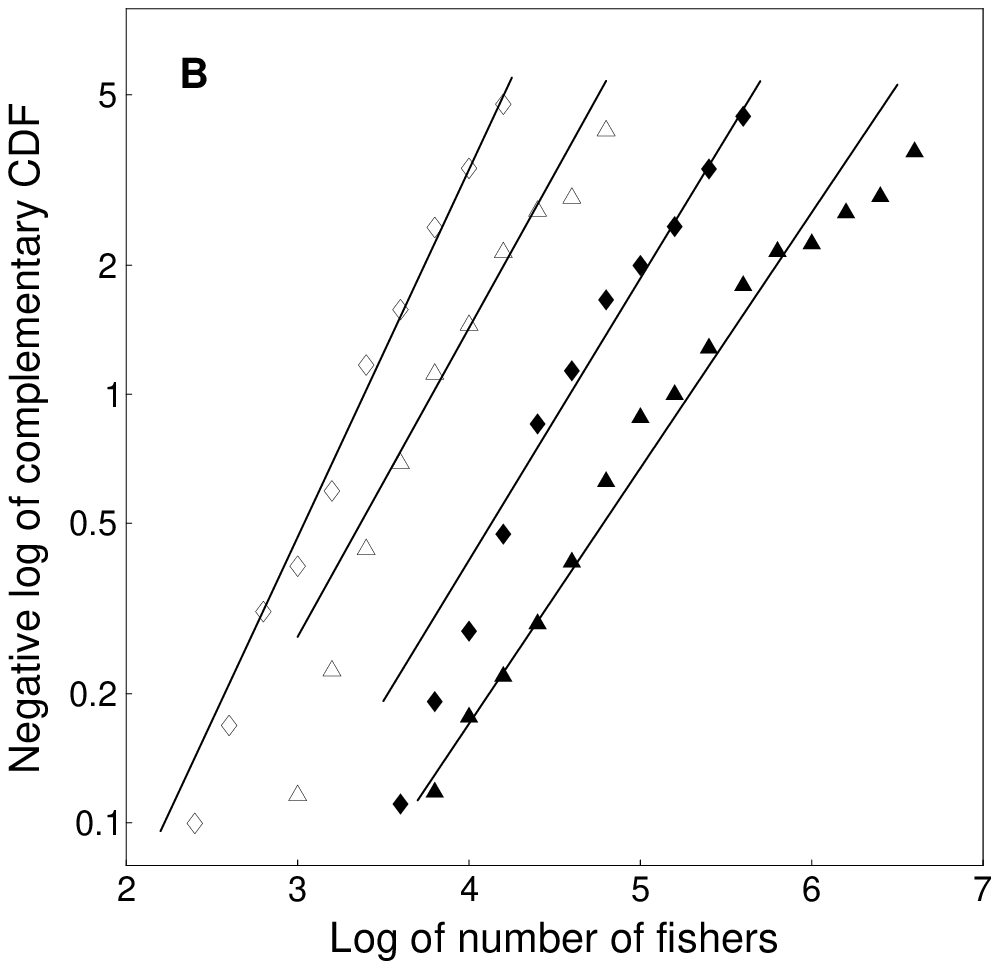}
}
 \caption{\small
}
 \label{fig:6}
\end{figure}
%

\begin{figure}[tb]
 \centering
{
 \includegraphics[width=6.5cm]{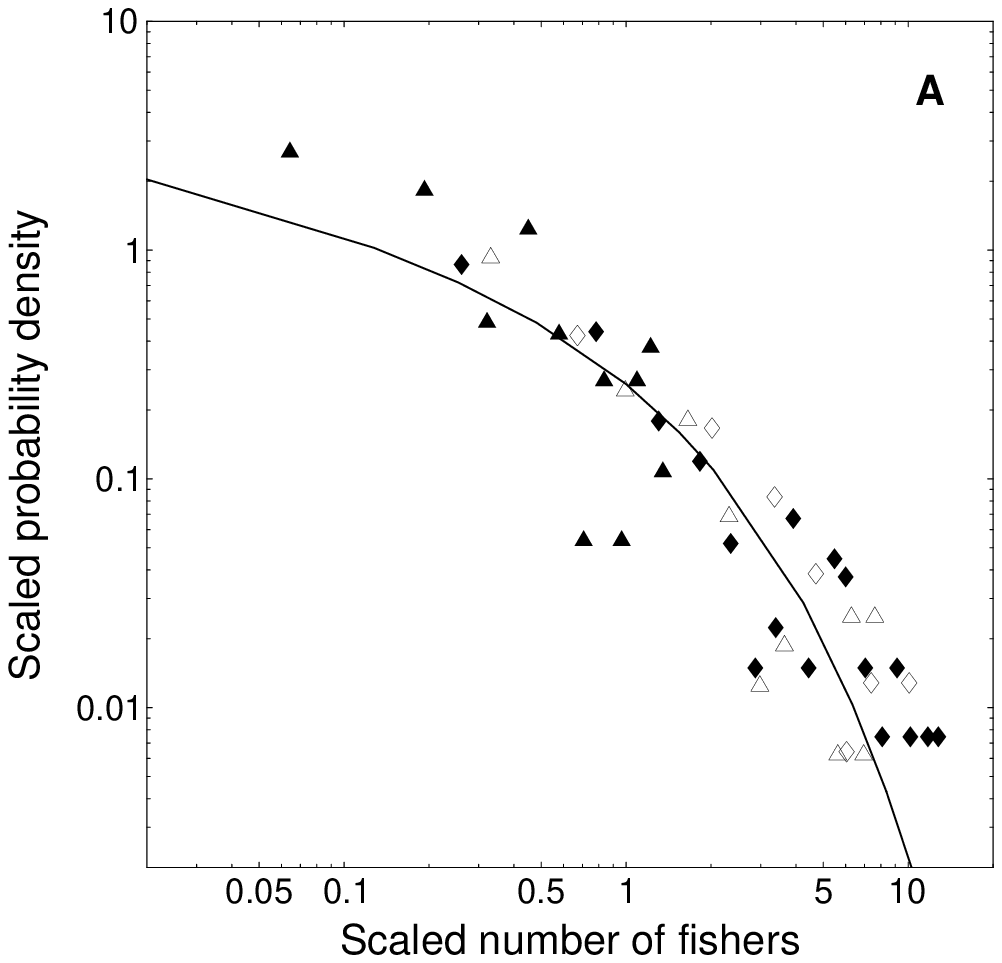}
 \hspace{0.5cm}
 \includegraphics[width=6.5cm]{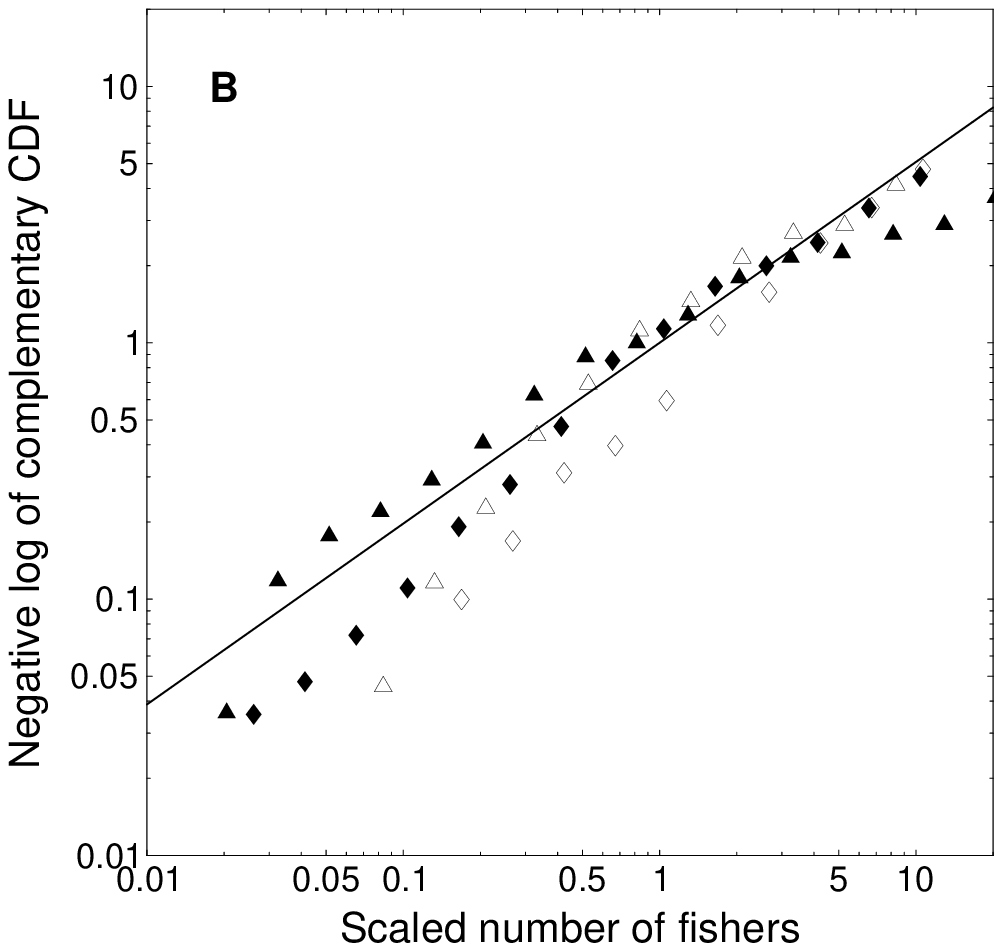}
}
 \caption{}
 \label{fig:7}
\end{figure}
%

\begin{figure}[tb]
 \centering
{
 \includegraphics[width=6.5cm]{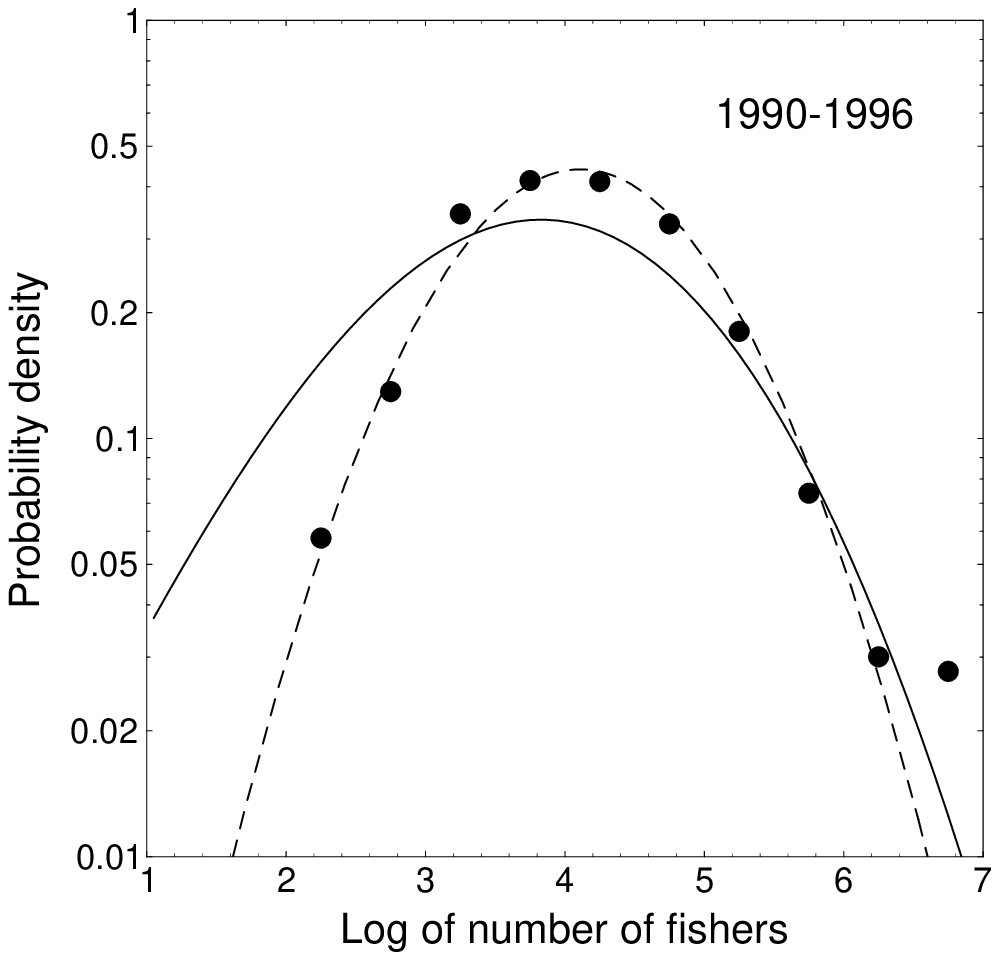}
}
 \caption{}
 \label{fig:8}
\end{figure}
%

\begin{figure}[p]
 \centering
{
 \includegraphics[width=6.5cm]{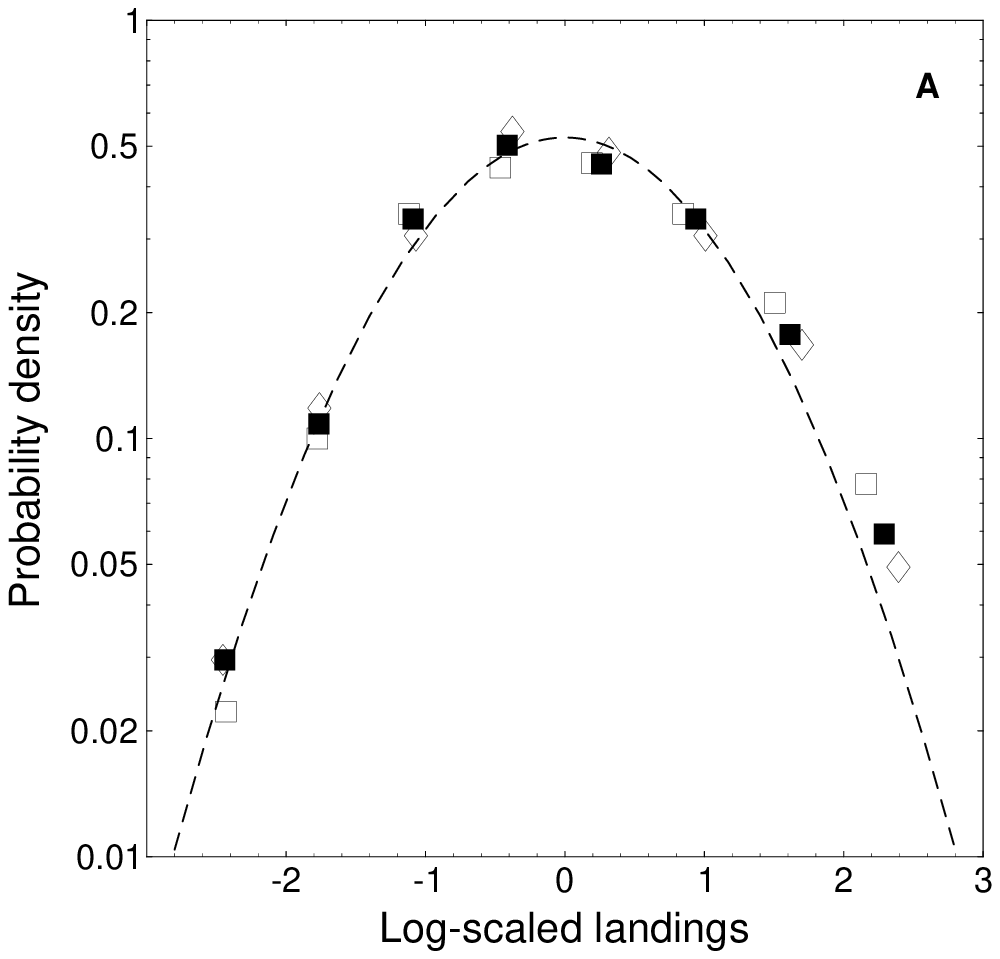}
 \hspace{0.5cm}
 \includegraphics[width=6.5cm]{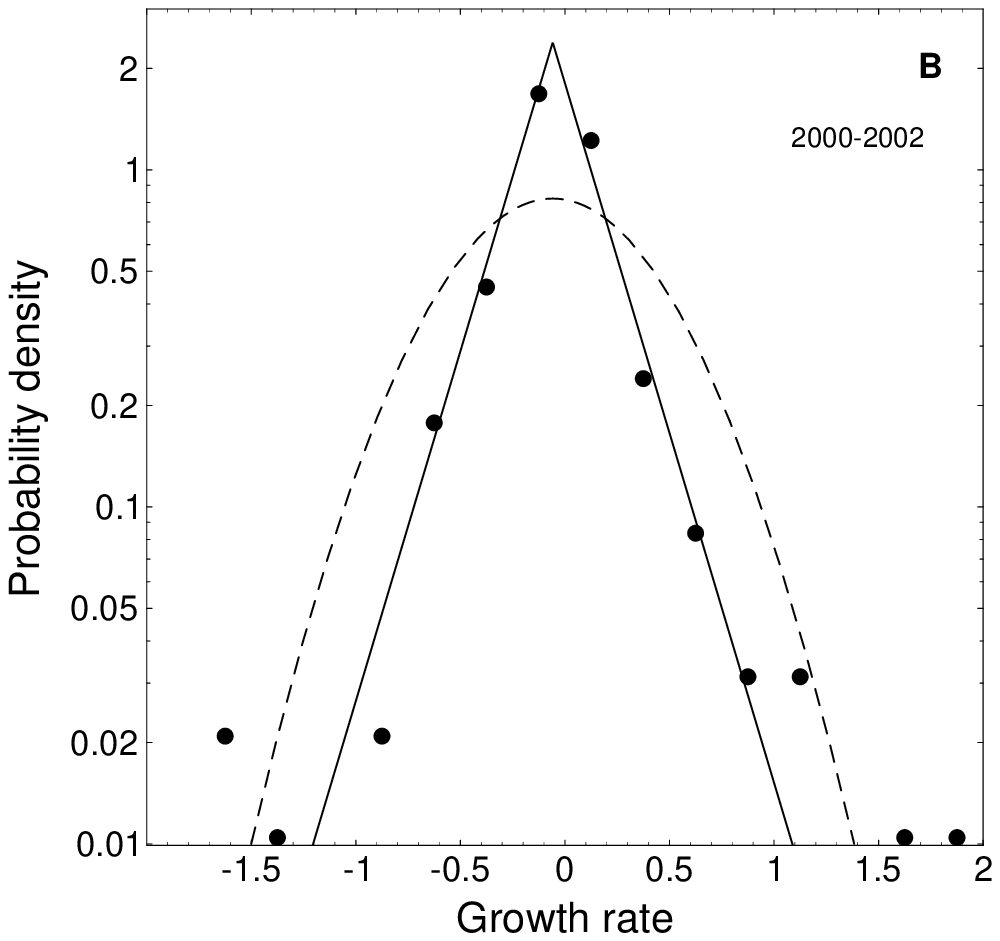}
}
 \caption{}
 \label{fig:9}
\end{figure}
\end{document}